\documentclass[12pt,prd,aps,amssymb,amsmath,tightenlines,showpacs]{revtex4}
\usepackage[utf8]{inputenc}
\usepackage{slashed}
\usepackage{graphicx,epsfig}
\usepackage{comment}
\usepackage{color}

\newcommand{\be}{\begin{equation}}
\newcommand{\ee}{\end{equation}}
\newcommand{\bea}{\begin{eqnarray}}
\newcommand{\eea}{\end{eqnarray}}

\begin{document}
\graphicspath{{FIGURE/}}
\topmargin=0.0cm

\title{Impact of New Physics on the EW vacuum stability
\\ in a curved spacetime background}

\author{E. Bentivegna$^{a,b}$}\email{eloisa.bentivegna@ct.infn.it}
\author{V. Branchina$^{a,b}$}\email{branchina@ct.infn.it}
\author{F. Contino$^{a,c}$}\email{contino.filippo92@gmail.com}
\author{D. Zappalà$^b$}\email{dario.zappala@ct.infn.it}

\affiliation{$^a$ Department of Physics and Astronomy,  
University of Catania, Via Santa Sofia 64, 95123 Catania, Italy}

\affiliation{$^b$ INFN, Sezione di Catania, Via Santa Sofia 64, 
95123 Catania, Italy}

\affiliation{$^c$ Scuola Superiore di Catania, 
Via Valdisavoia 9, 95123 Catania, Italy}

\affiliation{~~}

\affiliation{~~}

\date{\today}

\begin{abstract}
It has been recently shown that, contrary to an 
intuitive decoupling argument, the presence of new physics 
at very large energy scales (say around the Planck scale)
can have a strong impact on the electroweak vacuum lifetime.
In particular, the vacuum could be totally destabilized.
This study was performed in a flat spacetime background, and
it is important to extend the analysis to curved spacetime 
since these are Planckian-physics effects. It is generally 
expected that under these extreme conditions gravity 
should totally quench the formation of true vacuum bubbles, 
thus washing out the destabilizing effect of new physics. 
In this work we extend the analysis to curved spacetime  
and show that, although gravity pushes toward stabilization, 
the destabilizing effect of new physics is still (by far) 
the dominating one. In order to 
get model independent results, high energy new physics is
parametrized in two different independent ways: 
as higher order operators in the Higgs field, or 
introducing new particles with very large masses. 
The destabilizing effect is observed in both cases, 
hinting at a general mechanism that does not depend on 
the parametrization details for new physics, thus maintaining 
the results obtained from the analysis performed in flat 
spacetime. 
\end{abstract}

 

\maketitle  

\newpage 

\section{Introduction}

One of the most important goals of present theoretical and 
experimental particle physics is the search for New Physics 
(NP) beyond the Standard Model (BSM), even though direct 
experimental searches up to now have not revealed 
any sign of it. When looking for patterns towards BSM theories, 
the stability analysis\,\cite{cab,sher,lindn,sherrep,zagla,jones,sher2,
quiro,alta,isido,Espinosa:2007qp} of the electroweak (EW) 
vacuum plays a crucial role. 
Earlier studies were mainly focused on establishing  
bounds for the Higgs boson mass, based either on the requirement 
that the Higgs effective potential $V(\phi)$ could not 
take values lower than the EW minimum $v$ (so that the latter 
is the stable vacuum of the theory) or on the possibility 
that our Universe sits on a metastable (false) vacuum state
(i.e. $V(v)$ is not the absolute minimum of $V(\phi)$), 
with a lifetime larger than its own 
age\,\cite{turner},\cite{rees},\cite{sher},\cite{isido}. 
 
The discovery of the Higgs boson boosted new interest on 
the stability problem. Clearly the goal is no 
longer to derive bounds on its mass, but rather to perform 
more refined analyses that should allow to discriminate 
between absolute stability or metastability for the EW 
vacuum\cite{isidue, millington1, grinstein, millington2,
matschwartz}, 
to study the cosmological impact of the 
vacuum stability condition during and after 
inflation\,\cite{raja1, khan, raja2, kearney, goldberg, 
macdonald, ema1, ema2, okada, urbanowski1, urbanowski2}, 
and to test the impact that different NP scenarios can have 
on the vacuum stability 
condition\,\cite{NNLO,isiuno,isidue,bu,our1,our2,our3,haba,
ferreira,our4,chaka1}. 

On the theoretical side, the stability analysis has its roots in  
a pioneering work of Bender and collaborators\,\cite{bender}, where 
the tunneling for a quantum mechanical system with several degrees 
of freedom was studied with the help of saddle point techniques. 
This was later extended to quantum field theory by Coleman 
and Callan, who studied the decay of the false vacuum in a flat 
spacetime background\,\cite{coleman}, and then by Coleman and De 
Luccia\,\cite{cdl}, who included gravity in their analysis. 

Physically the false vacuum decay is triggered by quantum 
fluctuations, that induce a finite probability for a bubble of 
true vacuum to materialize in a false vacuum sea.
Both in flat and curved spacetime backgrounds, Coleman and 
collaborators considered a scalar theory where the potential 
$V(\phi)$ has a relative and an absolute minimum, at 
$\phi_{\rm false}$ and 
$\phi_{\rm true}$ respectively, such that the  
energy density difference $V(\phi_{\rm false}) - 
V(\phi_{\rm true})$ 
is much smaller than the height of the ``potential barrier''
$V(\phi_{\rm top}) - V(\phi_{\rm false})$, where 
$V(\phi_{\rm top})$ is the 
maximum of the potential between the two minima. 
Under these conditions  
the true vacuum bubble is separated from the false vacuum sea 
by a ``thin wall'', and this allows to treat the problem
analytically, within the so called ``thin wall'' approximation.

Going back to the SM, it is known that due to the top 
loop corrections the Higgs potential $V(\phi)$ bends down for 
values of $\phi >\, v $, where $v \sim 246$ GeV 
is the location of the EW minimum, and for the present 
experimental values of $M_H$ and $M_t$, namely $M_H \sim 125.09$ GeV 
and $M_t \sim 173.34$ GeV \,\cite{higgsmass,ATLAS:2014wva}, it 
develops 
a second minimum, much deeper than the EW one and at a much larger 
value of the field, $\phi_{\rm true}\, \gg \, 
v = \phi_{\rm false} $. 
The {\it instability scale} of the Higgs potential is then 
identified as the value \,$\phi_{\rm inst}$\, of the field 
such that $V(\phi_{\rm inst})=V(v)$ and $V(\phi) < V(v)$ for 
$\phi > \phi_{\rm inst}$. For the values of the Higgs and top masses 
reported above, it turns out that $\phi_{\rm inst} 
\sim 10^{11} {\rm GeV}$. 
Clearly the conditions under which the thin 
wall approximation can be applied are not fulfilled in the SM case, 
so the results of \cite{coleman} and \cite{cdl} cannot 
be directly applied.

The EW vacuum stability condition was first studied in a flat 
spacetime background, and the interesting 
possibility that the SM is valid all the way up to the Planck scale 
$M_P$, or more generally to some very large scale $M_{\rm large}$, 
meaning that NP  shows up only at this scale, was investigated.
In such a scenario,  naturally prompted by the lack of direct 
observation of hints of new physics, the analysis was performed 
under the assumption that the presence of NP  at 
$M_{\rm large}$ could be neglected for the computation of the tunneling 
time $\tau$ from the false to the true vacuum of the SM, so that 
$\tau$ was calculated by considering SM interactions 
only\,\cite{isido,Espinosa:2007qp,NNLO,isiuno,isidue,bu}. 
In fact it was argued that the relevant scale for tunneling is 
the instability scale $\phi_{\rm inst} \sim 10^{11}$ GeV, and that 
the contribution to the tunneling rate coming from NP  
that lives at the scale $M_{\rm large}$ should be suppressed. 
In other words, as $\phi_{\rm inst}  \sim 10^{11} 
{\rm GeV} \ll M_{\rm large}$, 
a decoupling effect was  expected\,\cite{Espinosa:2007qp}. 

It was later realized that the assumption that NP lives 
at $M_{\rm large}$ ($\gg \phi_{\rm inst}$) does not 
imply that it cannot 
affect the stability condition of the EW vacuum. On the contrary, 
the latter turns out to be  
very sensitive to unknown NP  even if it lives at scales 
far away from $\phi_{\rm inst}$, and the 
expected decoupling phenomenon does not take 
place\,\cite{our1,our2,our3,our4}.

The reason why the decoupling theorem does not hold in this case 
is that tunneling is a non-perturbative phenomenon\,\cite{our4},  
while the former applies when calculating scattering amplitudes 
in perturbation theory at energies $E$ much lower than 
$M_{\rm large}$. In this case the contributions to scattering
amplitudes from physics that lives at $M_{\rm large}$ is 
suppressed by factors of $E/M_{\rm large}$ to the appropriate 
power, and this is how physics at the scale $M_{\rm large}$ 
is decoupled from physics at the scale $E$.

For our tunneling phenomenon however, the bulk of the contribution 
to $\tau$ comes from the exponential of the (Euclidean) action 
calculated at the saddle point  of the path integral for the
tunneling rate, the so called {\it bounce solution} to the 
(Euclidean) Euler-Lagrange equation\,\cite{coleman}, and for 
this tree level contribution no suppression factors of the kind 
\,$(E/M_{\rm large})^{n}$\, can ever appear. 
If the Higgs  potential is modified by the presence of NP  
at $M_{\rm large}$, the new bounce is certainly different  from the 
one obtained when these terms are neglected. The action calculated 
for this new bounce solution is also modified and (once
exponentiated) it can give rise to a value of $\tau$ enormously
different from the one obtained when the NP terms are neglected.

The inclusion of gravity in the vacuum stability analysis was 
pioneered in\,\cite{cdl}, where the case of the thin wall regime 
was studied. 
For the transition from a false Minkowski vacuum to a true Anti-de 
Sitter (AdS) vacuum, it was shown that, when the size of  the 
Schwarzschild radius of the true vacuum bubble is 
much smaller than its size, i.e. when gravitational effects are 
weak, the probability of materialization of such a bubble is close 
to the flat spacetime result, while when the Schwarzschild radius 
becomes comparable to the bubble size, i.e.~in a strong 
gravitational regime, the presence of gravity stabilizes the 
false vacuum, preventing the materialization of a true 
vacuum bubble. In other words, gravity tends to stabilize the 
false vacuum, and in a strong  gravity regime the materialization 
of bubbles of true vacuum is quenched. 

The above results are obtained in the thin wall regime, but they 
are commonly considered of more general validity. In particular 
it was recently claimed\,\cite{espinosa,ridolfi,espinosa2} that 
in the presence of Planckian physics (strong gravity regime) 
the new bounce solutions\,\cite{our1,our2,our3,our4} to the 
(Euclidean) equation of motion modified by the presence of new 
physics at $M_{\rm large}$ should disappear, and that as 
a consequence 
these NP  terms could not induce any modification to the 
tunneling rate. 
 
As noted above however, the SM case is very far from  the thin 
wall regime analyzed in \,\cite{cdl}, and before jumping 
to any conclusion, and also prior to studies that consider  
the inclusion of new physics, the stability analysis for the 
SM in the presence of gravity has to be performed. 
An early attempt to study the impact of gravity on the EW vacuum  
decay rate was done in\,\cite{rych}, where a perturbative expansion 
of the bounce around the flat spacetime solution was considered.  
As later noted in\,\cite{our5} and then in\,\cite{rajantie} however,  
the boundary conditions for the bounce solution, that 
are essential in the calculation of the decay rate, are not 
respected already at the first order of the expansion, and  
the results of this work are at least questionable.   

In order to get close to the SM case, but still keeping a simple 
model as in\,\cite{cdl}, a scalar theory with a potential whose 
parameters can be adjusted to explore cases far from the thin wall 
regime was considered in\,\cite{our5}, and a numerical analysis of 
the false vacuum stability condition was performed. 
The main result is that for the potential that well approximates the  
SM case, the stabilizing effect of gravity 
is hardly seen even in very strong gravity  regimes. 
As suggested in\,\cite{banks,bousso,weinb}, the total quenching of 
the vacuum decay rate can eventually be reached at some very high 
scale. As shown in\,\cite{our5} however, for the SM case such 
an effect takes place in a far transplanckian regime where the 
theory has already lost its validity (similarly to what happens 
for the Landau pole in QED, where the latter occurs at such a high 
energy scale that the theory has lost its significance several 
orders of magnitudes below that scale).
The results obtained with the simple model considered 
in\,\cite{our5} were later confirmed in\,\cite{rajantie},  
where a bona fide SM Higgs effective potential was used.

The issue raised in\,\cite{espinosa,ridolfi,espinosa2} however,
namely the possibility that strong gravity effects should make 
the presence of Planckian NP harmless in the calculation 
of the tunneling time, is a crucial open question. In order 
to complete the stability analysis of the EW vacuum
it is of 
the greatest importance to understand if  gravity really cancels 
the destabilizing effect induced by NP at high energies, as 
claimed in\,\cite{espinosa,ridolfi,espinosa2}. 

In the present work we address these issues, that are very 
important for current studies and for model 
building of BSM physics, where we are often confronted with 
the possibility of considering NP  at Planckian 
and/or trans-Planckian scales. Anticipating on 
the results of the following sections, we will see that 
the tunneling time from the false to the true vacuum is still
strongly dependent on NP  even if it lives at very high 
($\gg \phi_{\rm inst}$) scales, thus confirming the results 
of the analysis performed in the 
flat spacetime background. 

The rest of the paper is organized as follows. In Section II
the general theoretical set-up for our work, mainly 
consisting of the equations that will be used 
in the subsequent numerical analysis, is presented. 
Moreover, in order 
to keep the present paper as self-contained as possible,
and also to check our tools against known results, the EW vacuum 
stability analysis in the absence of new 
physics in both flat and curved spacetime backgrounds
is briefly sketched, and the known results are
recovered. In section III we study the impact of NP  
at the Planck scale on the stability condition of the EW 
vacuum when the presence of gravity is taken into account,
parametrizing NP  in terms
of higher order operators. In Section IV a different
parametrization for NP  at high energy scales
is used, namely we introduce a new boson and a new fermion 
(with very large masses) coupled to the Higgs boson.
Section V is for our conclusions. 

\section{Theoretical background} 
\label{sec:thback}

In the present section we briefly review the theoretical background 
for the computation of the tunneling decay rate from a Minkowski 
false vacuum (minimum of the potential with vanishing energy density) 
to an Anti-De Sitter (AdS) true vacuum (minimum of the 
potential with negative energy density), considering both the 
flat and the curved spacetime background cases. 

{\it Flat spacetime}. 
Let us begin by considering the flat spacetime 
Euclidean action for a single component real scalar field $\phi$:
\begin{equation}\label{maction}
S[\phi]=\int d^4x \left [ \frac 1 2 (\partial_\mu \phi)^2 
+ V(\phi) \right ]\,,
\end{equation}
where $V(\phi)$ is a potential with a local minimum 
(\emph{false vacuum}) at $\phi=\phi_{\rm fv}$, and 
an absolute minimum (\emph{true vacuum}) at $\phi=\phi_{\rm tv}$. 

In order to calculate the false vacuum lifetime we have to look
for the so called \emph{bounce solution} to the 
Euclidean Euler-Lagrange equation that have
$O(4)$ symmetry and satisfy certain boundary 
conditions\,\cite{coleman}. If $r$ is the radial coordinate, 
the equation takes the form: 
\begin{equation}\label{meq}
\ddot \phi(r) + \frac{3}{r} \dot \phi (r) = \frac{d V}{d \phi}\,,
\end{equation}
where the dot indicates derivative with respect to $r$, and the 
boundary condition are:
\begin{equation}\label{mbc}
\phi(\infty)=0 \qquad \dot \phi(0)=0\,.
\end{equation}

Denoting with $\phi_b(r)$ the bounce solution, the 
action at $\phi_b$ is:
\begin{equation}\label{action}
S[\phi_b]=2 \pi^2 \int_0^\infty d r \ r^3 
\left [ \frac 1 2 \dot \phi_b^2 + V(\phi_b) \right ]\,,
\end{equation}
and the decay rate $\Gamma$ of the false vacuum is 
given by: 
\be \label{gamma}
\Gamma = D e^{-(S[\phi_b]-S[\phi_{\rm fv}])} \equiv D\, e^{-B}
\ee
where $B\equiv S[\phi_b]-S[\phi_{\rm fv}]$ is the so called 
{\it tunneling exponent}
and the exponential of\, $-B$ \, gives the ``tree-level'' 
contribution to the decay rate, while $D$ 
is the quantum fluctuation determinant. If $V(\phi_{\rm fv})=0$, 
the action $S[\phi_{\rm fv}]$ vanishes, 
and the tunneling exponent is simply $B=S[\phi_b]$. 
In order to determine the false vacuum decay rate in the 
flat spacetime case, in the following we integrate numerically 
Eq.\,(\ref{meq}) with boundary conditions (\ref{mbc}), and 
use (\ref{action}) to get the tunneling exponent $B$ of 
(\ref{gamma}). 

{\it Curved spacetime}. 
The next step is to study the impact of gravity on the 
vacuum decay rate, and to this end we consider the previous 
theory in a curved spacetime background. 
Including the Einstein-Hilbert term, the Euclidean 
action becomes:
\begin{equation}\label{gaction}
S[\phi, g_{\mu\nu}]=\int d^4 x \sqrt{g} 
\left [ -\frac{R}{16 \pi G} + \frac 1 2 g^{\mu \nu} 
\partial_\mu \phi \ \partial_\nu \phi + V(\phi) \right ]
\end{equation}
where $R$ is the Ricci scalar and $G$ is the Newton constant. 
Requiring again $O(4)$ symmetry, 
the (Euclidean) metric takes the form: 
\begin{equation}\label{metric}
ds^2= dr^2 + \rho^2(r) d\Omega_3^2
\end{equation}
where $d\Omega_3^2$ is the unit 3-sphere line element
and $\rho(r)$ is the volume radius of the 3-sphere at 
fixed $r$ coordinate\,\cite{cdl}. The bounce configuration needed 
to calculate the false vacuum transition rate is now given by
the field and the metric solution, $\phi_{b}(r)$ and 
$\rho_{b}(r)$ respectively, of the coupled equations 
($\kappa\equiv 8 \pi G$):
\begin{equation}\label{gequation}
\ddot \phi + 3 \ \frac{\dot \rho}{\rho} \ \dot \phi 
= \frac{d V}{d \phi} \qquad \dot \rho^2=
1+\frac{\kappa \rho^2}{3} 
\left ( \frac 1 2 \dot \phi^2 - V(\phi) \right )\,,
\end{equation}
where the first equation replaces (\ref{meq}), while the 
second is the only Einstein equation left by the symmetry. 
For the decay of a Minkowski false vacuum to a true AdS 
vacuum, the case of interest to us, the boundary conditions are:
\begin{equation} \label{bcond}
\phi_{_b}(\infty)=0 \qquad \dot \phi_{_b}(0)=0 
\qquad \rho_{_b}(0)=0\,.
\end{equation}

Asymptotically ($r \to \infty$) the bounce $\phi_b(r)$ 
approaches the constant false vacuum solution $\phi_{\rm fv}$, where 
$V(\phi_{\rm fv})=0$ (Minkowski vacuum). In the same limit, 
from\,(\ref{gequation})$_{_2}$ we see that the bounce solution 
metric $\rho_b(r)$ approaches the flat spacetime metric: 
\begin{equation}\label{as}
\rho_b(r)= r+c\,.
\end{equation}
In the thin wall regime the constant $c$ is obtained 
analytically\,\cite{cdl}, while in general it is determined from 
the numerical integration of (\ref{gequation}).

Differentiating\,(\ref{gequation})$_{_2}$ with respect to $r$ 
we get:
\be \label{xx2}
\ddot \rho=-\frac{\kappa}{3} \rho \left 
( \dot \phi^2 + V( \phi) \right )\,,
\ee
that is a useful equation that we will use in our analysis as 
it is more robust than (\ref{gequation})$_{_2}$ for numerical 
integration\,\cite{goto}. 
Finally the Ricci scalar $R$ in terms of $\rho$
is given by:
\be \label{curvature}
R=-\frac{6}{\rho^2} \left ( \rho \ddot \rho + \dot 
\rho^2 - 1 \right )
\,.
\ee

Inserting the above results in (\ref{gaction}), the action for 
the bounce $(\phi_{_b},\rho_{_b})$ becomes: 
\begin{equation}\label{gravact}
S_b\equiv S[\phi_b,\rho_b]=2\pi^2\int_{0}^{\infty} 
dr \left \{ \rho_{b}^3 \left [ \frac 1 2 \dot \phi_{b}^2
+V(\phi_{b}) \right ] - \frac 3 \kappa (\rho_{b} 
\dot \rho_{b}^2 + \rho_{b}) \right \} + 
\frac{6 \pi^2}{\kappa}(\rho_{b}^2 \dot \rho_{b})\Big |_{0}^{\infty}\,.
\end{equation}

As the false vacuum action $S[\phi_{\rm fv},\rho_{\rm fv}]$ contains 
the same boundary term (the last one) of (\ref{gravact}), in the 
tunneling exponent $B$ this term does not appear. Neglecting it, 
and using (\ref{gequation}) in (\ref{gravact}) we finally get:

\begin{equation}\label{azionegrav}
S_b = 4\pi^2 \int_{0}^{\infty} 
d r \left [ \rho_b^3 V(\phi_b) - \frac 3 \kappa \rho_b \right ]\,.
\end{equation}

From (\ref{azionegrav}), the false vacuum action 
$S_{\rm fv} \equiv S[\phi_{\rm fv},\rho_{\rm fv}]$ can be easily derived.
As already said, we 
are interested in the decay of a Minkowski false vacuum 
(the state that corresponds to the minimum of the potential with 
vanishing energy density, $V(\phi_{\rm fv})=0$) to an AdS true 
vacuum. For the false vacuum case, $\phi(r)=\phi_{\rm fv}$, and 
$\dot \rho^2=1$, i.e. $\rho_{\rm fv}(r)= r + const$.
This latter constant is fixed and is \,$c$\, of Eq.\,(\ref{as}), 
as the bounce metric asymptotically tends to the flat spacetime 
metric of the Minkowski false vacuum. In calculating the action 
$S_{\rm fv}$ we must then integrate over the Minkowski spacetime 
from a real radius equal to zero up to infinity. This corresponds 
to the integration range $[-c,\infty[$ in the $r$ coordinate,
\be
S_{\rm fv}=-4\pi^2 \int_{-c}^{\infty} d r \ \frac 3 \kappa (r+c)=-4\pi^2 \int_{0}^{\infty} d r \ \frac 3 \kappa (r+c)-\frac{6\pi^2}
{\kappa}c^2\,,
\ee
so that for the tunneling exponent $B=S_b-S_{\rm fv}$ we have:
\be \label{gB}
B=4\pi^2 \int_{0}^{\infty} dr \left [ \rho^3 V(\phi_b) 
- \frac 3 \kappa (\rho-r-c) \right ]+\frac{6\pi^2}{\kappa}c^2\,.
\ee

In the following, the decay rate of the Minkowski false vacuum 
into the AdS vacuum will be calculated by integrating numerically 
Eqs.\,(\ref{gequation})$_{_1}$ and (\ref{xx2}) with boundary 
conditions\,(\ref{bcond}). Then with the help of  
(\ref{gB}) the tunneling exponent will be obtained. 

As we are interested in the stability analysis of the EW 
vacuum, in our case the scalar field $\phi$ is the Higgs 
field, and the potential $V(\phi)$ is the Higgs effective 
potential. More specifically it is the renormalization 
group improved potential, that can be 
written as:
\be \label{potential}
V_{\rm SM}(\phi) \sim \frac 1 4 \lambda_{\rm SM} (\phi)\phi^4\,,
\ee
where $\lambda_{\rm SM}(\phi)$ is the quartic running coupling 
$\lambda_{\rm SM}(\mu)$ ($\mu$ is the running scale) with 
$\mu=\phi$\,\,\cite{CW,sher}.

In order to get $\lambda_{\rm SM}(\mu)$, the system of 
RG equation of the SM couplings has to be run. 
In\,\cite{rajantie}, where a stability analysis of the 
SM in the presence of gravity was performed, an improved 
Higgs potential obtained with the help of three-loops
beta functions was used, while in previous analyses 
two-loop beta functions were considered\,\cite{isidue}. 
In this respect, we note that the consistent 
counting of loops in the beta functions of different SM 
particles poses a problem in itself, as it was 
found that taking the same order for all the different 
components of the SM (Yukawa, gauge and quartic couplings) 
actually leads to inconsistencies\,\cite{sannino}. 

The purpose of the present work however is to study 
the impact that NP  at high energies can
have on the stability condition of the EW vacuum when the  
SM coupling to gravity is taken into account. We 
are then not interested in precision measurements 
and/or refinements of previous analyses. For our 
illustrative scopes, the differences between these 
cases (the two loops counting, the three 
loops counting, and the ``consistent'' counting proposed 
in\,\cite{sannino}) are minimal, and they have no impact 
on the results and conclusions of our analysis. 

We can then leave aside these questions and work in 
a simplified yet very robust framework, by using 
a good approximation of the SM effective potential 
that was obtained in\,\cite{moss} by fitting the 
two-loops improved Higgs potential 
with the three parameter function\,\cite{moss}: 
\be\label{lam}
\lambda_{\rm SM}(\phi)=\lambda_* + \alpha \left ( \ln \frac{\phi}{M_P} 
\right )^2 + \beta \left ( \ln \frac{\phi}{M_P} \right )^4\,,
\ee
where $M_P=1/\sqrt G$ is the Planck mass. The fit gives: 
\be\label{p}
\lambda_*=-0.013 \qquad \alpha=1.4 \times 10^{-5} 
\qquad \beta=6.3 \times 10^{-8}\,.
\ee
In the following we work with the Higgs potential 
(\ref{potential}) with $\lambda_{\rm SM}(\phi)$ given by 
(\ref{lam}) and (\ref{p}).  

Both in the flat and curved spacetime cases, an 
important parameter is the size $\cal R$ of 
the bounce, defined as the value of $r$ such that

\be
\phi_b({\cal R})=\frac 1 2 \phi_b(0)\,.
\ee

Going back to (\ref{gamma}) for the vacuum decay
rate, we note that a good approximation to the prefactor 
for the case that we are considering is given in terms of 
the bounce size ${\cal R}$ and of $T_U$, the age of the 
Universe, and the EW vacuum 
tunneling time $\tau=\Gamma^{-1}$ turns out to be\,\cite{our3}:
\be \label{tau}
\tau \simeq \left (\frac{{\cal R}^4}{T_U^3} \right) e^B\,.
\ee
In the following we use (\ref{tau}) to calculate the false vacuum 
lifetime.

Before ending this section and moving to the study of the 
impact of NP on the EW vacuum stability, we 
would like to test our tools starting with the known cases of 
the flat and curved spacetime backgrounds 
in the absence of NP  (i.e.\,considering the SM alone), 
and briefly sketch the analysis for these cases. 

{\it Flat spacetime}.
In order to proceed with the numerical solution of the 
bounce equation (\ref{meq}), we begin by scaling the 
dimensionful field $\phi$ and the radial coordinate $r$
to dimensionless quantities, $x$ and $\varphi (x)$ respectively, 
by using Planck units: 
\be \label{scaling}
x \equiv M_P r \qquad \varphi(x) \equiv \frac{\phi(r)}{M_P}
\ee
Eq.(\ref{meq}), the boundary conditions (\ref{mbc}) and
the potential (\ref{potential}) then become:
\be \label{smeq}
\varphi''(x) + \frac 3 x 
\varphi'(x)=\frac{dU}{d \varphi}
\ee
\begin{equation}\label{dimlesseq}
\varphi(\infty)=0 \qquad \varphi'(0)=0
\end{equation}
\begin{equation}\label{potenzialescalato}
U(\varphi)=\frac 1 4 \varphi^4 \left 
( \lambda_* + \alpha \ln^2 \varphi + \beta \ln^4 \varphi \right )\,,
\end{equation}
where the prime indicates the derivative respect to $x$. 
After the rescaling (\ref{scaling}), the tunneling exponent 
(\ref{action}) becomes:
\be \label{te}
B=2 \pi^2 \int_0^{\infty}dx \ x^3 \left [ \frac 1 2 
\varphi'_b(x)+U(\varphi_b) \right ]\,.
\ee

Solving numerically the bounce equation (\ref{smeq}), 
with the Higgs potential given by 
(\ref{potenzialescalato}) and (\ref{p}), 
and inserting the result for $\varphi_b(x)$ in 
(\ref{te}), after 
using the values found for $B$ and ${\cal R}$, namely 
$B=2025.27$ and  ${\cal R} = 10.7597$, 
we finally get for the lifetime $\tau$ of the EW vacuum: 
\be \label{tf}
\tau_{\rm flat} \sim 10^{639} T_U\,,
\ee
in very good agreement with the results known 
in the literature. This is the first test of our numerical 
method, and also shows that we are considering a good 
approximation for the Higgs potential. \newline
 
{\it Curved spacetime}. 
As in the case of flat spacetime, we move to dimensionless 
quantities. Defining the dimensionless curvature 
$a(x)=M_P\, \rho(r)$, Eqs.\,(\ref{gequation})$_{_1}$ 
and (\ref{xx2}) (that will be used in 
the following numerical integration) become:
\begin{equation}\label{seqgrav}
\varphi'' + 3 \ \frac{a'}{a} \ \varphi'=\frac{d U}{d \varphi} 
\qquad \qquad  a'' = - \frac{8\pi}{3}a 
\left (\varphi'^{\,2} + U \right )\,,
\end{equation}
where the potential $U(\varphi)$ is the same as in 
(\ref{potenzialescalato}). The corresponding 
boundary conditions are:
\begin{equation} \label{sgbc}
\varphi(\infty)=0 \qquad \varphi'(0)=0 \qquad a(0)=0 
\qquad a'(0)=1 \,.
\end{equation}

As we have already said, $\rho(r) \sim r$ for $r \to \infty$, and  
the asymptotic ($x \to \infty$) behavior of the bounce 
solution in the presence of gravity is the same as in the 
flat spacetime case. 

In terms of dimensionless quantities, from (\ref{gB}) we find 
for tunneling exponent:
\be \label{sgB}
B=4 \pi^2 \int_0^{\infty} dx \left [ a_b^3 U(\varphi_b)-
\frac{3}{8\pi}(a_b-x-c) \right ]+\frac{3 \pi}{4}c^2
\ee
where $c$ (as already said above) is the constant that determines 
the asymptotic behavior of the metric $a(x)$ for $x\to\infty$,
while ($\varphi_b$, $a_b$) is the bounce solution to 
the system (\ref{seqgrav}).

\begin{figure} 
	\begin{minipage}[b]{7cm}
		\centering
		\includegraphics[width=1.1\textwidth]{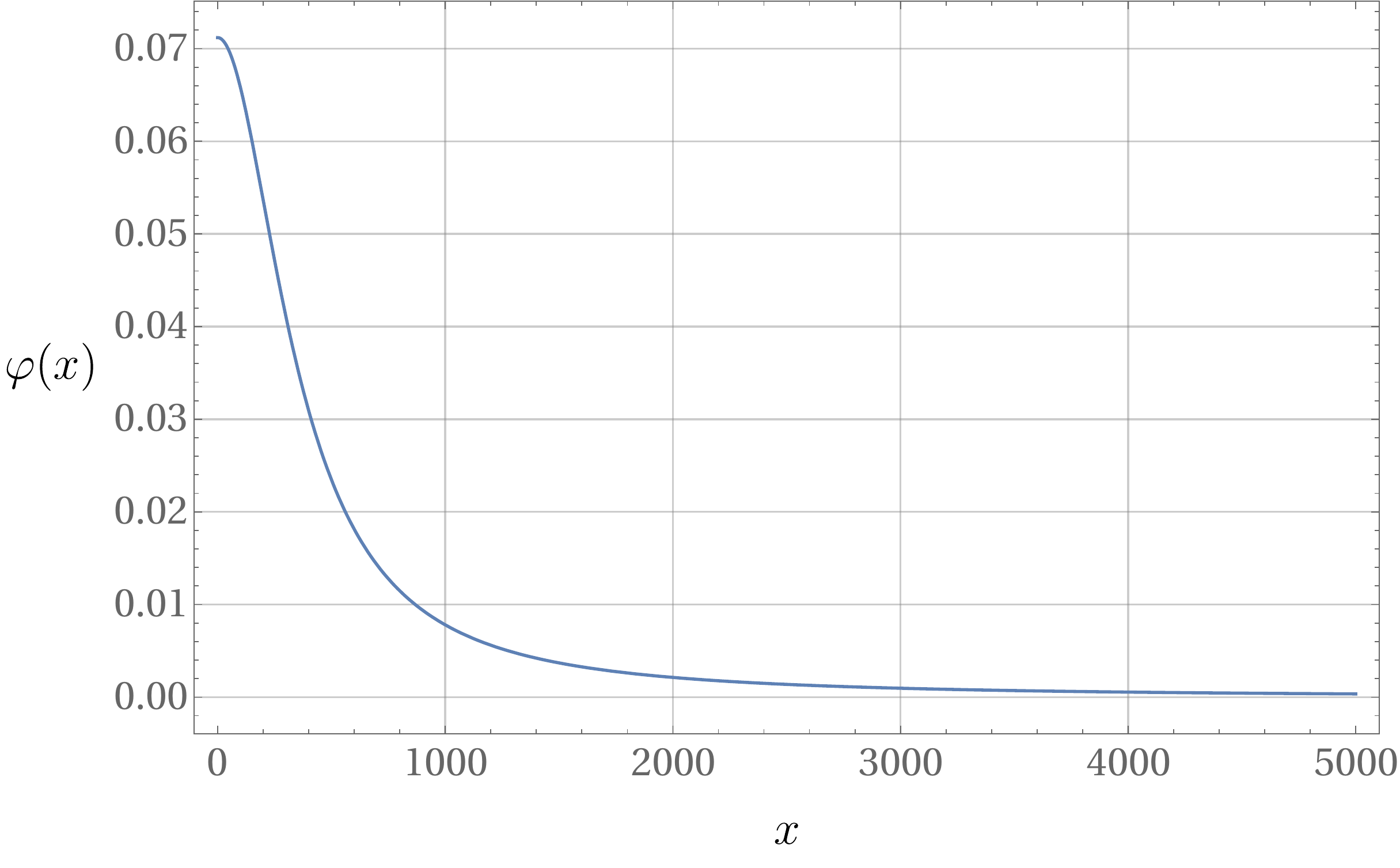}
	\end{minipage}
	\hspace{5mm}
	\begin{minipage}[b]{7cm}
		\centering
		\includegraphics[width=1.1\textwidth]{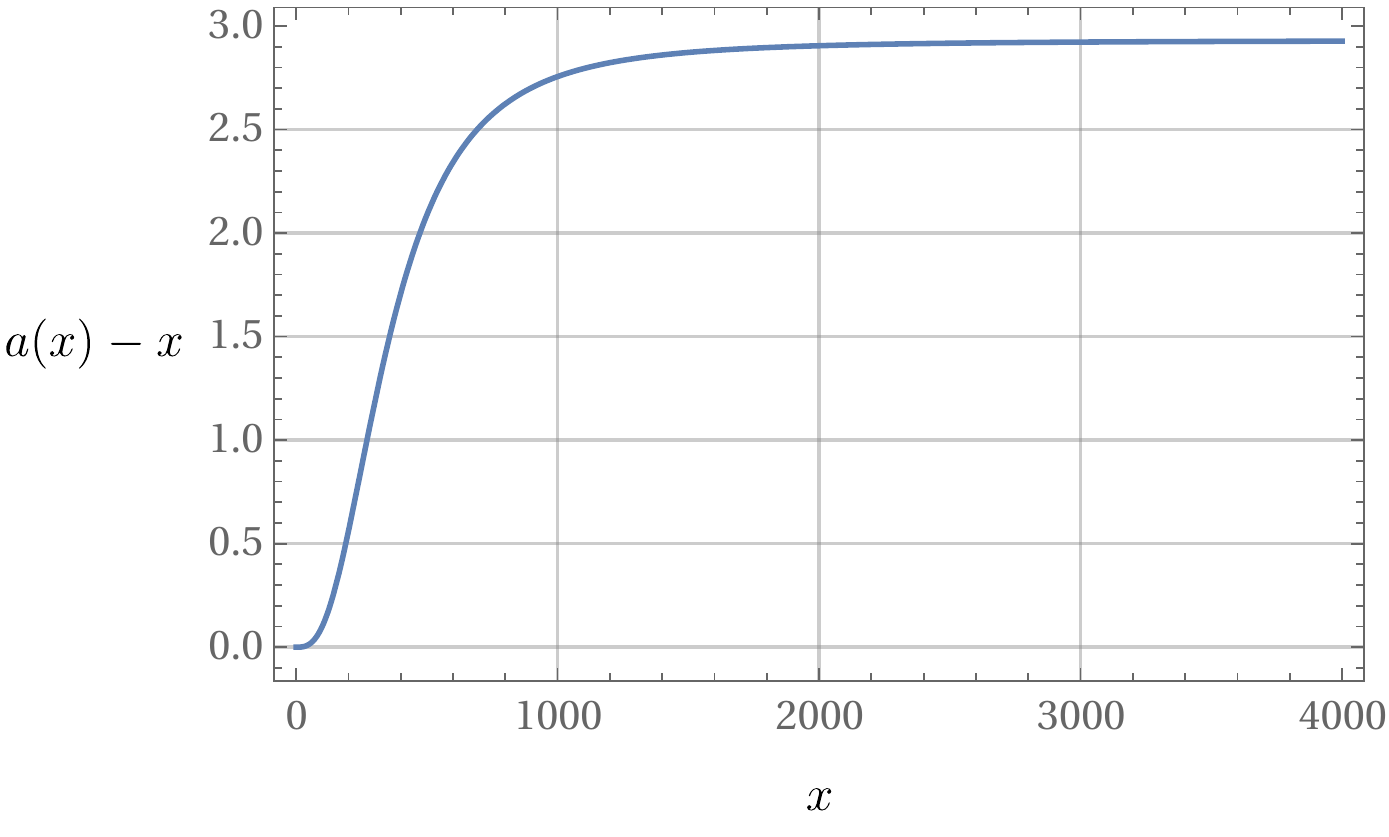}
	\end{minipage}
	\caption{{\it Left panel}: Profile of the bounce 
solution $\varphi(x)$ in the presence of gravity. It is
obtained for the potential (\ref{potenzialescalato}) with 
the parameters 
$\lambda_*$, $\alpha$, $\beta$ given in\,(\ref{p}).  
The center of the bounce is at 
$\varphi(0)=0.0712$, its size is $\mathcal R =350.2996$ and 
the tunneling exponent is $B=2062.5836$. {\it Right panel}: 
Difference between the curvature radius and its asymptotic 
value, $a(x)-x$, for the same parameters as in the left panel.}
\label{fig:figgrav}
\end{figure}

In the left panel of Fig.\,\ref{fig:figgrav} the bounce profile
$\varphi_b(x)$ is plotted. The right panel shows the difference 
$a_b(x)-x$: we clearly see how asymptotically $a_b(x)$ 
reaches the Minkowskian behavior $a(x) \sim x + c$, and we 
can read the value of the constant $c$. 
Finally, with the help of (\ref{tau}), 
we obtain the tunneling time in the presence of gravity:
\be \label{gt}
\tau_{\rm grav} \sim 10^{661} T_U\,.
\ee

Once again we observe that the above result is in good agreement 
with known results\,\cite{rajantie}. Moreover, comparing 
(\ref{gt}) with the corresponding flat spacetime tunneling 
time (\ref{tf}), we see that gravity (as expected) tends 
to stabilize the EW vacuum. 

\begin{figure}[!h]
	\includegraphics[width=0.8\textwidth]{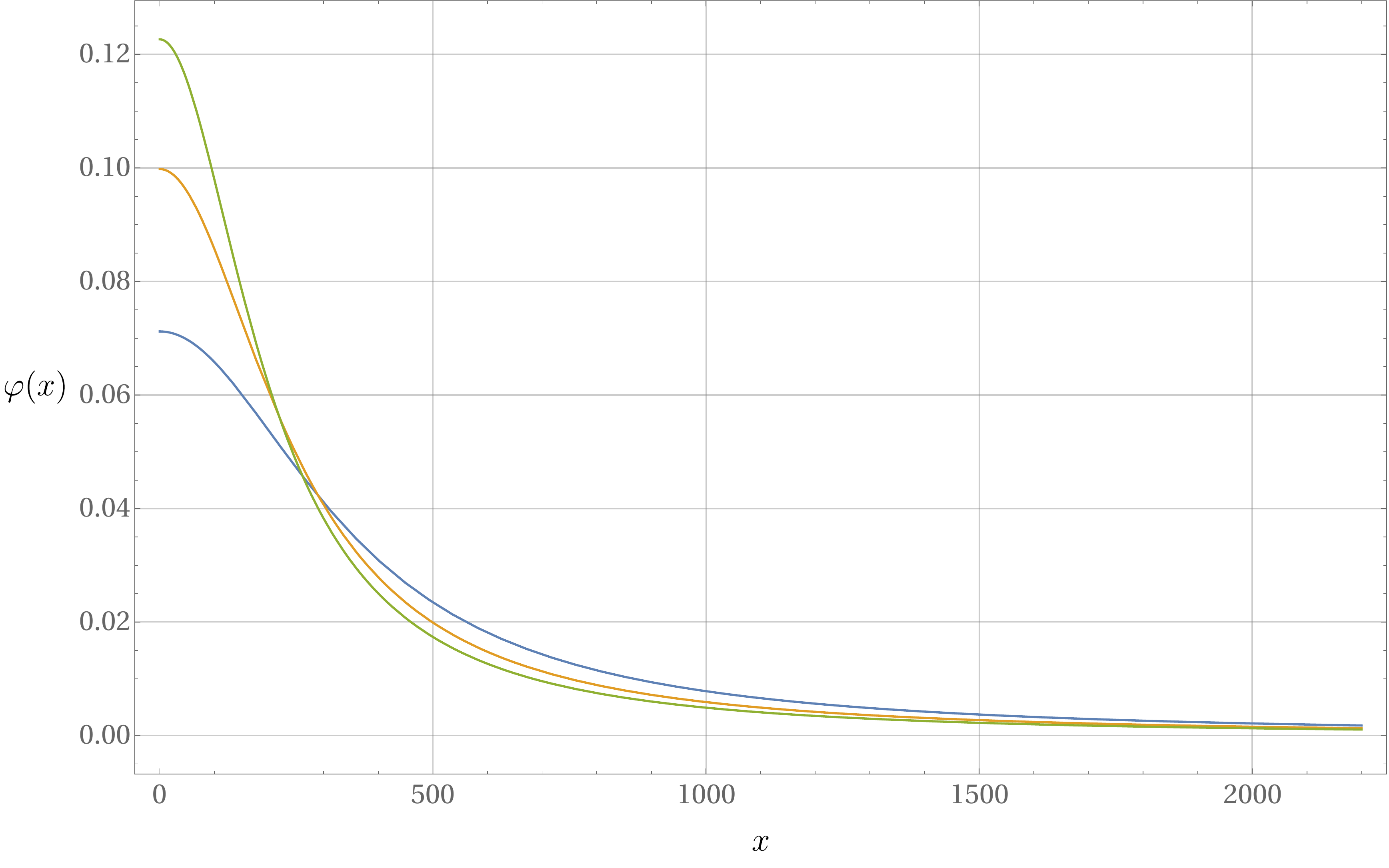}
	\caption{The blue curve is the profile of the 
bounce solution obtained for the potential (\ref{newp}) with 
$\lambda_6=0$ and $\lambda_8=0$, i.e. in the absence of new 
physics. The yellow curve is the profile 
of the bounce solution for $\lambda_6=-0.3$ and $\lambda_8=0.3$,
while the green one is the profile of the bounce obtained for 
$\lambda_6=-0.4$ and $\lambda_8=0.4$. Note that with increasing 
values of the couplings the center of the bounce $\varphi(0)$ 
becomes larger while the size diminishes.}
	\label{fig:blambda}
\end{figure}

\section{New physics: Higher order operators}

The results briefly presented in Section II are known and 
concern the stability analysis under the assumption 
that, in the event that unknown NP lives at a scale 
$M_{\rm large}$ much greater than the instability scale 
$\phi_{\rm inst}$, it has no impact on the stability condition 
of the EW vacuum. In other words it is assumed that new physics
at $M_{\rm large}$ is decoupled from the physics that triggers 
the EW vacuum decay, and that it should be possible to 
calculate the tunneling rate ignoring these terms.  

The analysis of the previous section is essential to set 
the proper framework where the effects of the presence 
of NP  at $M_{\rm large}$ can be properly investigated. 
To be specific, we consider the case where NP  lives 
at the Planck scale $M_P$ (i.e. we choose $M_{\rm large}=M_P$) 
and parametrize it as in\,\cite{our1,our2,our3}
with the help of higher powers of $\phi$ added to the Higgs 
potential:
\be \label{pnp}
V_{\rm NP}(\phi)=\frac{\lambda_6}{6} 
\frac{\phi^6}{M_P^2}+\frac{\lambda_8}{8} \frac{\phi^8}{M_P^4}\,.
\ee

It was shown in\,\cite{our1,our2,our3} for the flat spacetime 
case that when $\lambda_6<0$ and $\lambda_8>0$ the potential 
(\ref{pnp}) destabilizes the EW vacuum. In other words, 
these NP  terms favor the nucleation of true vacuum 
bubbles and, depending on the specific values of $\lambda_6$ 
and $\lambda_8$, this destabilization effect could 
dramatically reduce the EW vacuum lifetime $\tau$ 
in\,(\ref{tf}) and make it even shorter than the age 
of the Universe $T_U$. We now consider the same kind of 
analysis in the presence of gravity. 

Adding the NP  terms (\ref{pnp}) to the SM Higgs potential 
(\ref{potential}), and moving again to dimensionless 
quantities, the new dimensionless potential $U(\varphi)$ 
becomes: 
\be \label{newp}
U(\varphi)=\frac 1 4 \varphi^4 
\left ( \lambda_*+\alpha \ln^2 \varphi+\beta\ln^4 \varphi 
+ \frac 2 3 \lambda_6 \varphi^2 + \frac 1 2 
\lambda_8 \varphi^4 \right )\,.
\ee

We are now ready to study the impact of high energy 
NP  on the EW vacuum stability condition in the 
presence of gravity. 

A first important result of our analysis is that for 
each value of the couple ($\lambda_6$, $\lambda_8$) there is a 
different bounce 
solution to Eqs.\,(\ref{seqgrav}), all of them being 
different from the solution obtained for the SM alone, 
i.e. the case $\lambda_6=0$, $\lambda_8=0$. 

Contrary to the expectations of 
\,\cite{espinosa,ridolfi,espinosa2}
then, gravity does not induce the disappearance 
of the ``new'' bounce solutions related to the presence of new 
physics, here parametrized in terms a given couple 
($\lambda_6$, $\lambda_8$). In other words, 
gravity does not wash out the destabilizing effect induced 
by Planckian new physics: the decoupling of Planckian physics 
from the tunneling phenomenon does not take place. 

In order to illustrate these results, in Fig.\,\ref{fig:blambda} 
we show bounce solutions to Eqs.\,(\ref{seqgrav})
for $\lambda_6=-0.3$,  
$\lambda_8=0.3$ (yellow curve), $\lambda_6=-0.4$, 
$\lambda_8=0.4$ (green curve) and compare them with the 
corresponding $\lambda_6=0$, $\lambda_8=0$ (blue curve) 
case. The profiles 
obtained are definitely {\it new solutions} to these equations  
related to the specific values of $\lambda_6$ and $\lambda_8$,
clearly different from the bounce (blue curve) obtained for 
the SM alone ($\lambda_6=0$ and $\lambda_8=0$). 

With the help of\, (\ref{tau}) we now calculate the 
EW vacuum lifetime for different values of the NP  
couplings $\lambda_6$ and $\lambda_8$. The fourth 
column of Table\,\ref{tab:lambda}  contains different 
values of the tunneling time obtained for different couples 
($\lambda_6$, $\lambda_8$). For comparison, the third column 
contains the corresponding values of $\tau$ for the flat
spacetime analysis. First of all we note that the effect 
already seen in the previous section (also reported 
in the first line of the table, the case $\lambda_6=0$, 
$\lambda_8=0$), namely that the presence of 
gravity tends to stabilize the EW vacuum, is maintained 
even in the presence of new physics. 

\begin{table}
	\centering
	\begin{tabular}{c c | c c}
		$\lambda_6$ \hspace{3mm} & $\lambda_8$ \hspace{3mm} & \hspace{3mm} $\tau_{\rm flat}/T_U$ \hspace{3mm} & 
$\tau_{\rm grav}/T_U$ \\ \hline
		
		$0$ \hspace{3mm} & $0$ \hspace{3mm} & \hspace{3mm} $10^{639}$ \hspace{3mm} & $10^{661}$ \\ \hline
		
		$-0.05$ \hspace{3mm} & $0.1$ \hspace{3mm} & \hspace{3mm} $10^{446}$ \hspace{3mm} & $10^{653}$ \\ \hline
		
		$-0.1$ \hspace{3mm} & $0.2$ \hspace{3mm} & \hspace{3mm} $10^{317}$ \hspace{3mm} & $10^{598}$ \\ \hline
		
	    $-0.15$ \hspace{3mm} & $0.25$ \hspace{3mm} & \hspace{3mm} $10^{186}$ \hspace{3mm} & $10^{512}$ \\ \hline
	    
	    $-0.3$ \hspace{3mm} & $0.3$ \hspace{3mm} & \hspace{3mm} $10^{-52}$ \hspace{3mm} & $10^{287}$ \\ \hline
	    
	    $-0.45$ \hspace{3mm} & $0.5$ \hspace{3mm} & \hspace{3mm} $10^{-93}$ \hspace{3mm} & $10^{173}$ \\ \hline
	    
	    $-0.7$ \hspace{3mm} & $0.6$ \hspace{3mm} & \hspace{3mm} $10^{-162}$ \hspace{3mm} & $10^{47}$ \\ \hline
	    
	    $-1.2$ \hspace{3mm} & $1.0$ \hspace{3mm} & \hspace{3mm} $10^{-195}$ \hspace{3mm} & $10^{-58}$ \\ \hline
	    
	    $-1.7$ \hspace{3mm} & $1.5$ \hspace{3mm} & \hspace{3mm} $10^{-206}$ \hspace{3mm} & $10^{-106}$ \\ \hline
	    
	    $-2.0$ \hspace{3mm} & $2.1$ \hspace{3mm} & \hspace{3mm} $10^{-206}$ \hspace{3mm} & $10^{-121}$ \\ \hline
	\end{tabular}
		    \caption{Tunneling time for 
different values of $\lambda_6$ and $\lambda_8$, both for 
the flat and curved spacetime cases. We note that 
although gravity tends to stabilize the EW vacuum (the 
tunneling time $\tau_{\rm grav}$ is always higher than 
the corresponding one in flat spacetime $\tau_{\rm flat}$),
new physics has always a strong impact.}
		    \label{tab:lambda}
\end{table}

However, a simple inspection of this table shows that even 
though the presence of gravity tends to 
stabilize the EW vacuum as compared to the corresponding 
flat spacetime case, still for $O(1)$ values of the new 
physics couplings $\lambda_6$ and $\lambda_8$ the tunneling
time can be made smaller than the age of the Universe $T_U$.
Let us consider just a couple of examples. For $\lambda_6=-0.3$ 
and $\lambda_8=0.3$ for instance, the EW vacuum in the flat 
spacetime background is unstable, being $\tau 
\sim 10^{-52} T_U$, but for the corresponding case with  
gravity included we observe a stabilization of the EW 
vacuum: $\tau \sim 10^{287} T_U$. There is a competition between 
the destabilizing effect of NP  and the stabilizing 
effect of gravity. In this example, gravity takes over new
physics and as a result the EW vacuum turns out to be stable.
However for larger (absolute) values of the NP  
couplings, the destabilizing effect of NP  takes over 
the stabilizing effect of gravity.
For instance, for $\lambda_6=-1.2$ and $\lambda_8=1.0$, 
despite the stabilizing effect of gravity 
($\tau_{\rm grav} \gg \tau_{\rm flat}$), the EW vacuum turns out to 
be unstable: $\tau_{\rm grav}  \sim 10^{-58} T_U$. 

\begin{figure}
\includegraphics[width=1.0\textwidth]{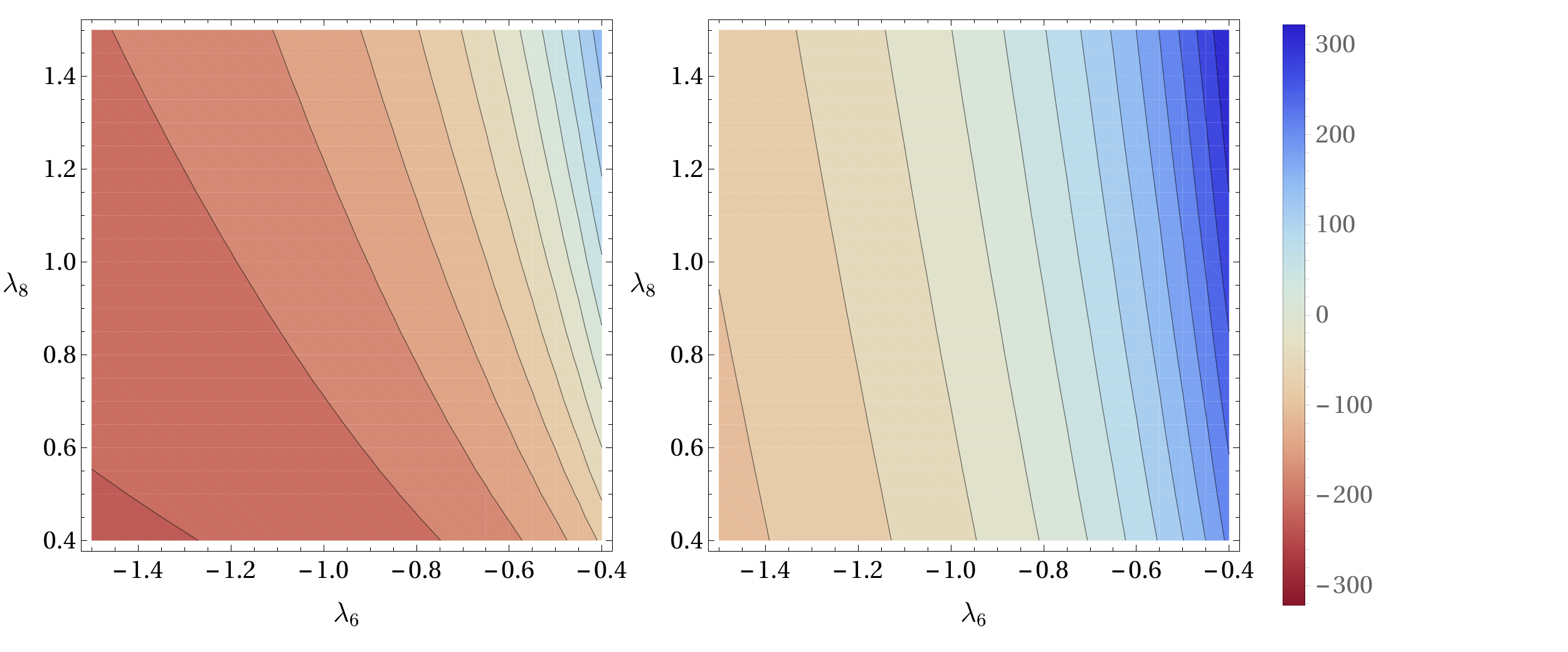}
	\caption{Stability diagrams in 
the ($\lambda_6, \lambda_8$) plane for $\log_{10} \tau$ with 
the potential of (\ref{potenzialescalato}) and parameters 
$\lambda_*$, $\alpha$, $\beta$ given in\,(\ref{p}).
{\it Left panel}: flat spacetime case.  
{\it Right panel}: curved spacetime case.
Colors in this figure are associated to the values of $\log_{10} \tau$
according to the scale shown on the right. 
Note that the inclusion of gravity induces a broad enlargement
of the region with stable EW vacuum ($\log_{10} \tau >0$).
}
\label{fig:liv}
\end{figure}

The results discussed above with the help of 
Table\, \ref{tab:lambda} are 
better summarized in Fig.\,\ref{fig:liv} where stability diagrams in 
the $(\lambda_6, \lambda_8)$ plane are presented for the range
of values $-1.5 < \lambda_6 < 0.4$ and $0.4 < \lambda_8 < 1.5$.
In the left panel the flat spacetime case is considered, and we
see that the stability region ($\tau > T_U$) is confined to the 
upper right corner of this figure. The right panel shows the  
stability diagram for the same range of values of the new physics
couplings, and we see that the stability range here takes 
half of the diagram. These figures illustrate the features 
discussed above. On the one hand gravity tends to stabilize the 
EW vacuum. On the other hand, for natural values of the 
NP  coupling constants $\lambda_6$ and $ \lambda_8$, 
i.e. for $O(1)$ values of these couplings, the destabilizing 
effect of Planckian NP  can take over the stabilizing 
effect of gravity.  

\begin{figure}
\includegraphics[width=0.5\textwidth]{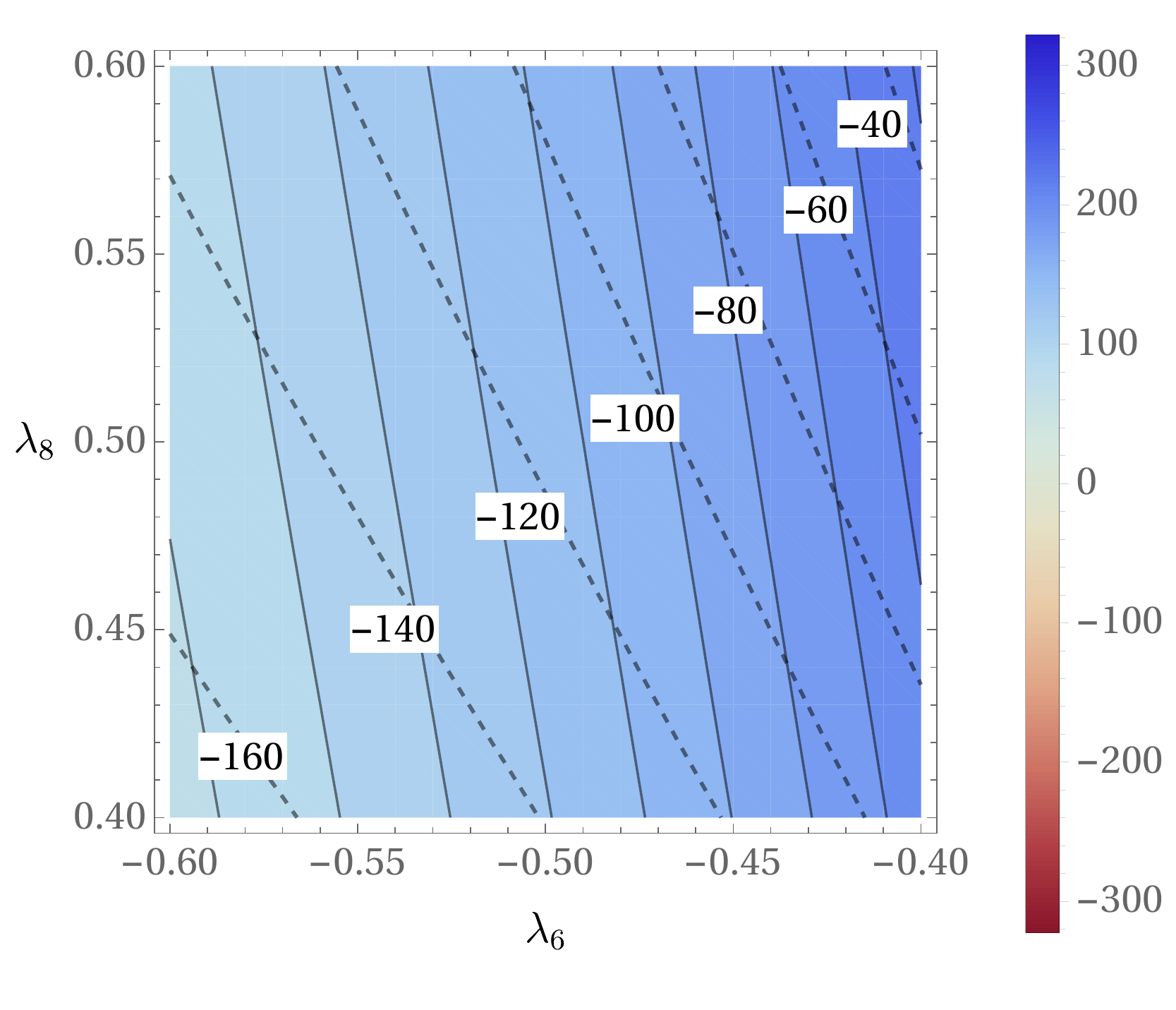}
	\caption{Zoom of part of the right panel of 
Fig.\,\ref{fig:liv}, with the inclusion of the flat spacetime 
level curves (dashed lines). For the considered range of 
$\lambda_6$ and $\lambda_8$, in the flat spacetime case 
the EW vacuum is always unstable, while in the presence of  
gravity it is always stable.}
\label{fig:diag}
\end{figure}

Fig.\,\ref{fig:diag} is a zoom of the stability diagram of 
Fig.\,\ref{fig:liv} for the 
case where gravity is included in the restricted range of values 
$-0.6< \lambda_6 < -0.4$ and $0.4 < \lambda_8 < 0.6$, where the 
dashed black lines are the level curves of the flat spacetime 
case (with the corresponding values of $\tau_{\rm flat}$ reported). 
For the whole range of values of $\lambda_6$ and $\lambda_8$
considered in this figure, gravity wins over the destabilizing 
effect of NP  ($\tau_{\rm grav}>T_U$) and the EW vacuum 
turns out to be stable, while in the flat spacetime background 
it turns out to be unstable ($\tau_{\rm flat} < T_U$). 

Let us summarize the results of the present section. On the one 
hand we have shown that the inclusion of gravity in the stability 
analysis of the EW vacuum does not cause the disappearance of 
the ``new'' bounce solutions found in the flat 
spacetime case\,\cite{our1,our2,our3} that in turn cause 
the destabilization of the EW vacuum to the point that its lifetime 
can become smaller than the age of the Universe. On the 
other hand we have seen that the stability condition of the EW 
vacuum is the result of a competition between the destabilizing 
effect of NP  and the stabilizing tendency of gravity. 
In particular we have also seen that outside a certain range of 
values of the NP  couplings $\lambda_6$ and $\lambda_8$, 
the destabilizing effect of Planckian NP  always takes 
over, thus making the EW vacuum unstable. 

\section{New physics: fermions and bosons with large masses}

In the present section the stability analysis of the EW vacuum
will be performed by considering a different parametrization 
for NP  at high energy scales.
Actually in\,\cite{our1,our2,our3} and in the previous section  
the analysis was performed by 
parametrizing NP  at the Planck scale in terms of few  
higher order (non-renormalizable) operators. 
This is just a convenient and efficient way of mimicking the 
presence of new physics, clearly not an (illegitimate) 
truncation of the UV completion of the SM. 
Some authors however expressed a certain skepticism
on these results, suggesting that this effect should disappear  
when the infinite tower higher dimensional operators of 
the renormalizable UV completion of the SM is taken into 
account, so that the expected decoupling of very   
high energy physics from the mechanism that triggers the 
decay of the false vacuum should be recovered. It was actually 
suspected that this effect takes place above the physical 
cutoff, where the control of the theory is lost\,\cite{espigiu}.

Although it is understandable that the parametrization of 
NP  in terms of higher order operators could be 
the source of a certain confusion, the destabilizing 
effect has nothing to do with this parametrization. 
For the case of a flat spacetime background 
in\,\cite{our4} the stability analysis was performed 
by parametrizing NP  in terms of renormalizable 
additional terms, with a fermion and a boson with very 
high masses that interact with the Higgs field, and it 
was shown that the destabilizing effect found 
in\,\cite{our1,our2,our3} is still present.  
 
In this section we present the same kind of analysis
of \cite{our4} taking into account the presence of gravity
(i.e.\,considering the case of a curved spacetime background), 
and show that as for the case of the parametrization used in 
the previous section, gravity does not produce any washing 
out of the destabilizing effect of new physics, although it 
slightly mitigates it.  

In order to illustrate the destabilization phenomenon we 
consider as in\,\cite{our4} a renormalizable model that is 
not a realistic high energy UV-completion of the SM but 
is very appropriate to the purposes of the present work. 
NP  lives at very high energy scales and is 
parametrized by adding to the SM a scalar field $S$ and a 
fermion field $\psi$ that interact in a simple way with the 
Higgs field $\phi$, with masses $M_S$ and $M_f$ of the scalar 
and fermion respectively well above the instability scale:
$M_S , M_f \gg \phi_{\rm inst}$. 

Apart from the kinetic terms, 
the additional terms in the Lagrangian are: 
\begin{eqnarray}\label{int}
\Delta \,{\cal L}&=&\frac{M_S^2}{2}S^2 
+ \frac{\lambda_S}{4} S^4 + g_S \phi^2 S^2 
+M_f \bar\psi\psi 
+g_f \phi \bar \psi \psi\,. 
\end{eqnarray}  

To understand how a NP  Lagrangian of this 
kind can arise in a physical setup, we note that the large mass 
term $M_f$ can be thought as a 
sort of heavy right handed ``neutrino'' in the framework of 
a see-saw mechanism. While the corresponding light ``neutrino'' 
is totally harmless for the stability of the EW vacuum, 
the heavy ``neutrino'' can play an important role in 
destabilizing the vacuum. The scalar field $S$ counterbalances 
the destabilizing effect of $\psi$. Note that models
with new scalar fields coupled to the Higgs (although 
admittedly unrealistic) have already been used to provide 
a stabilization mechanism for the Higgs effective 
potential\,\cite{hung,quiros}.

Before proceeding with our work, it is worth to mention 
a problem concerning the physical observables involved in the 
stability analysis, namely the gauge dependence of the effective 
potential away from the extrema\,\cite{sch1, diluzio, sch2}.
In particular, absolute stability bounds on the Higgs mass 
(formally gauge independent) turn out to be gauge dependent 
at any order of perturbation theory. Only when a consistent
resummation is considered the result can be made gauge 
independent, providing a slight improvement in mass 
bounds\,\cite{sch2}. The gauge dependence of the 
instability scale has also been investigated and the range 
of uncertainty identified\,\cite{diluzio}, but it is not 
known if it is possible to calculate this quantity in a 
gauge independent manner. Moreover, for the main quantity of 
interest to us, namely the false vacuum decay rate, again 
we know that it is a formally gauge-invariant quantity. 
In a truncated 
perturbative expansion, however, order-by-order gauge 
independence can possibly be achieved only after resumming 
the appropriate terms, as it was done for the energies 
at the minima of the effective potential\,\cite{sch1}. 
Having these warnings in mind, and waiting for improvements 
on these gauge dependence issues, we proceed now with the 
analysis of our model following the usual pattern.

For the purposes of the present work, it is sufficient to 
consider the impact of these additional terms on the Higgs 
effective potential $V(\phi)$ at the one-loop level only. 
In fact we do not need a better 
level of precision as we are not interested in extracting 
numbers but we only want to illustrate the destabilization 
effect that arises from very high energy physics (see 
also the considerations developed below Eq.\,(\ref{potential})).
The one-loop contribution to $V(\phi)$ from these terms is: 
\begin{eqnarray}\label{renormpot}
V_1(\phi)&=&
\frac{\left(M_S^2+2g_S\phi^2\right)^2}{64\pi^2}
\left[\ln\left(\frac{M_S^2+2g_S\phi^2}{M_S^2}\right)
-\frac{3}{2}\right]\nonumber\\
&-&\frac{\left(M_f^2+g_f^2\phi^2\right)^2}{16\pi^2}
\left[\ln\left(\frac{M_f^2+g_f^2\phi^2}{M_S^2}\right)
-\frac{3}{2}\right] \,,
\end{eqnarray}
where the renormalization scale $\mu$ is taken as 
$\mu=M_S$. In this respect we note that at very high 
values of the running scale the SM quartic coupling 
reaches a plateau: $\lambda_{\rm SM}(\mu)$ has 
practically the same value in the whole range 
$[M_f,M_S]$, and this is why we can use 
$\lambda_{\rm SM}(M_S)$ as the threshold value for the 
coupling (even though strictly speaking we should use 
$\lambda_{\rm SM}(M_f)$ (see below)), and can choose 
$\mu=M_S$ as the renormalization (threshold) scale.

The presence of the high energy NP  of \,(\ref{int}) 
is then taken into account by adding to the SM potential 
$V_{\rm SM}(\phi)$ in \,(\ref{potential}) and 
(\ref{lam}) the contribution coming from $V_1(\phi)$.
To this end we have to implement the matching conditions 
described below. First of all we expand the 
potential $V_1(\phi)$ in powers of $\phi$ and isolate
the constant, the $\phi^2$ and the $\phi^4$ terms. 
Then at the threshold scale $M_f$ we 
require that: (i) the renormalized cosmological 
constant $\Lambda$, given by the sum of all the constant 
terms (those coming from the SM potential and those coming 
from $V_1(\phi)$) vanishes, $\Lambda(\mu=M_f) \sim 0$; 
(ii) the renormalized mass term, given by the sum 
of all the coefficients of $\phi^2 $, and identified with
the SM mass parameter $m^2_{\rm SM}(\mu=M_f)$ 
at the scale $M_f$, vanishes: $m^2_{\rm SM}(\mu=M_f) \sim 0$ 
(more precisely we neglect this term to a very high degree of 
accuracy for the large values of $\phi$ considered); 
(iii) the renormalized quartic 
coupling, given by the sum of all the coefficients of 
$\phi^4$, is identified with the SM quartic coupling at 
the scale $M_f$, $\lambda_{\rm SM}(\mu=M_f)$. In other words, 
at the scale $M_f$ this coefficient is matched with the value 
of the quartic coupling obtained by considering the 
running of the renormalization group equations for the SM 
couplings alone. 

The above requirements for the renormalized 
cosmological constant and mass are well known features. 
For the renormalized $\Lambda$ (apart from the fine tuning 
problem) we can practically consider that  
$\Lambda(\mu=0) \sim \Lambda (\mu=M_f) \sim 0$. 
The same is true for the renormalized mass, for which 
we take $m^2(\mu=0) \sim m^2(\mu=M_f) \sim 0$,
meaning that we neglect the $\phi^2$ term as 
compared to the $\phi^4$ and other terms for these large 
values of $\phi$, and that the running of the renormalized 
mass is totally harmless in this respect. For the quartic 
coupling we have a true matching condition. In fact
we require that at the threshold scale $\mu=M_f$ the quartic 
coupling coincides with $\lambda_{\rm SM}(\mu=M_f)$, that is
obtained by running the renormalization 
group equations for the SM couplings only. 
Practically starting from the scale $M_f$, the potential 
is given by the SM contribution $V_{\rm SM}(\phi)$ plus 
the contribution of $V_1(\phi)$ subtracted of its 
constant, quadratic and quartic powers of $\phi$, that we 
call $\overline V_1(\phi)$ from now on: 
\be\label{Vtot}
V_{\rm tot}(\phi)=\frac{1}{4}\lambda_{\rm SM}(\phi)\phi^4+ 
{\overline V}_1(\phi)\,.
\ee

We are now ready to use our model of high energy 
NP  to calculate the EW vacuum lifetime for 
different values of the masses $M_f$ and $M_S$ of $\psi$ 
and $S$, and for different values of the coupling constants.
For our illustrative purposes we have chosen to consider the 
four following examples: 
(i) $M_S = 2.5\times 10^{-1}$, $M_f = 3\times 10^{-4}$, 
$g_S = 0.96$, $g_f^2 = 0.5$ ; 
(ii) $M_S = 2.0 \times 10^{-1}$, $M_f = 10^{-4}$,
$g_S = 0.9$, $g_f^2 = 0.5$ ; 
(iii) $M_S = 2.0 \times 10^{-1}$, $M_f = 10^{-3}$, 
$g_S = 0.95$, $g_f^2 = 0.4$ ; 
(iv) $M_S = 1.5 \times 10^{-1}$, $M_f = 5\times 10^{-3}$,
$g_S = 0.92$, $g_f^2 = 0.4$. 

\begin{figure} 
	\begin{minipage}[b]{7cm}
		\centering
		\includegraphics[width=1.1\textwidth]{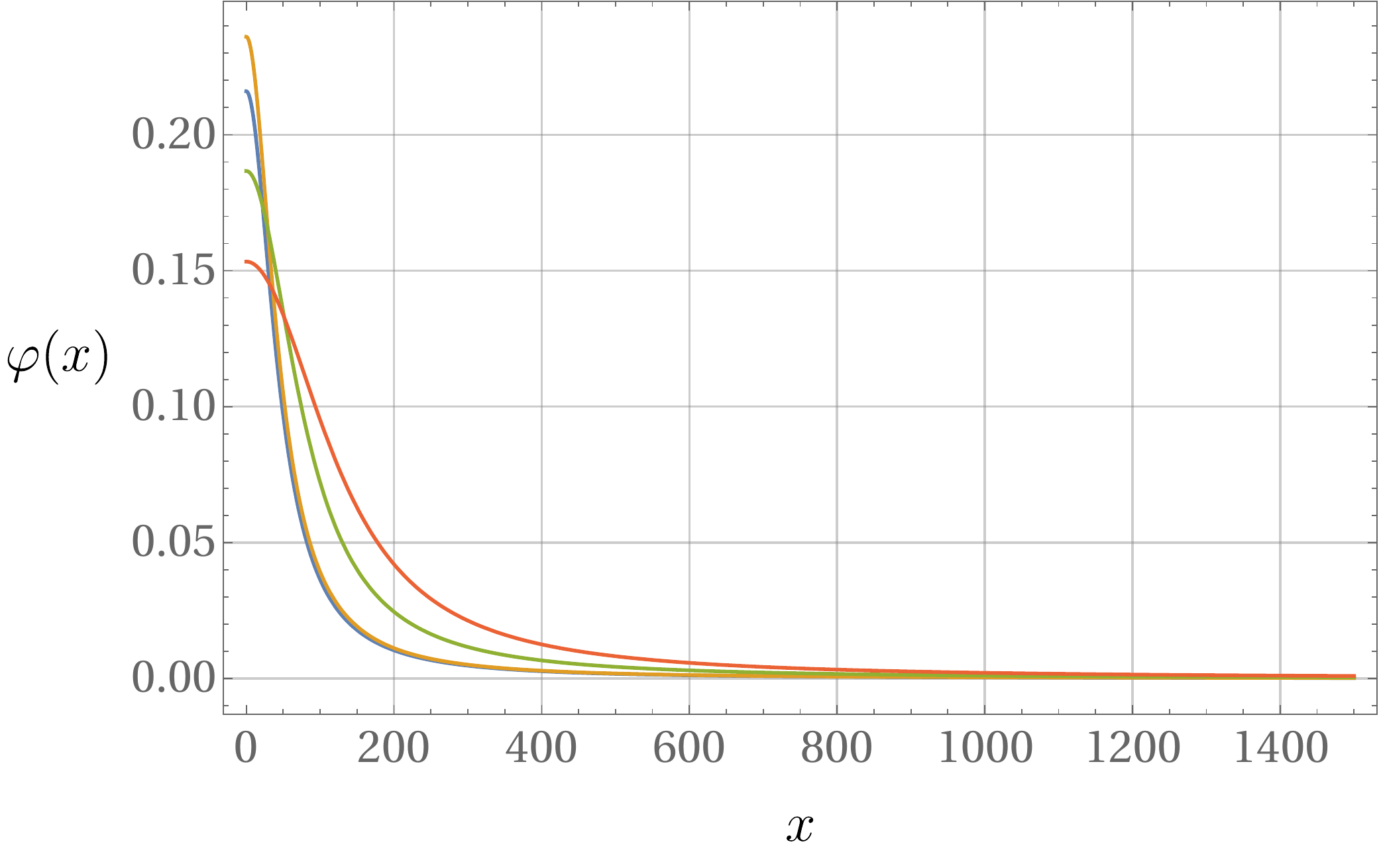}
	\end{minipage}
	\hspace{5mm}
	\begin{minipage}[b]{7cm}
		\centering
		\includegraphics[width=1.1\textwidth]
{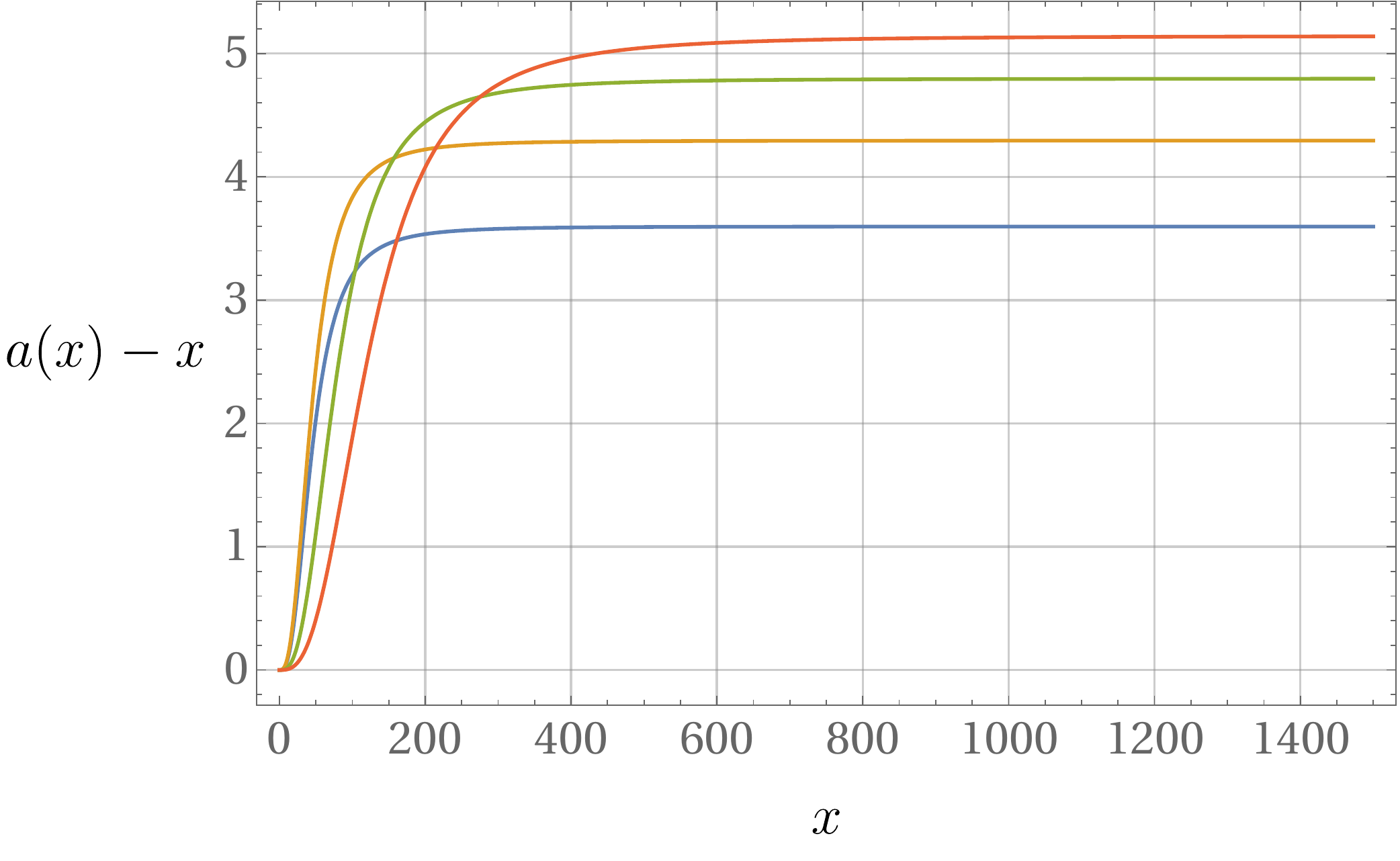}
	\end{minipage}
	\caption{{\it Left panel}: Profile of the bounce
solutions $\varphi(x)$ 
for the potential (\ref{Vtot}) 
relative to the four cases considered in the text:
$M_S = 2.5\times 10^{-1}$, $M_f = 3\times 10^{-4}$, 
$g_S = 0.96$, $g_f^2 = 0.5$ (yellow) ; 
$M_S = 2.0\times 10^{-1}$, $M_f = 10^{-4}$,
$g_S = 0.9$, $g_f^2 = 0.5$ (blue); 
$M_S = 2.0\times 10^{-1}$, $M_f = 10^{-3}$, 
$g_S = 0.95$, $g_f^2 = 0.4$ (green); 
$M_S = 1.5\times 10^{-1}$, $M_f = 5\times 10^{-3}$,
$g_S = 0.92$, $g_f^2 = 0.4$ (red). 
{\it Right panel}: the corresponding difference between the 
curvature radius and its asymptotic value, $a(x)-x$, for 
the same parameters as in the left panel.}
	\label{fig:NPren}
\end{figure}

First of all we have to solve the bounce equations 
(\ref{seqgrav}) for $\varphi(x)$ and $a(x)$. 
In Fig.\,\ref{fig:NPren}
the profiles of the bounce solutions $\varphi_{b}(x)$ for 
the four different cases (i), (ii), (iii) and (iv) 
and the corresponding plots of $a_b(x) - x$ 
are presented, with colors yellow, blue, green and red 
respectively. 

These are the first relevant results of the present section. 
They show that, contrary to the claims 
in\,\cite{espinosa,ridolfi,espinosa2}, the presence of gravity 
does not cause the disappearance  of the new bounce solutions 
due to the presence of high energy NP, that were already found 
in the case of the analysis carried in the flat spacetime 
background\,\cite{our4}. Therefore they confirm the results of 
the previous section, where
high energy NP was parametrized in terms of 
higher order operators, and once again show that the 
appearance of new bounce solutions is not an artifact of 
the specific parametrization used in Section III.   

Using (\ref{tau}) to calculate the vacuum lifetime, 
for the examples considered above we find in units of 
$T_U$ (going from (i) to (iv)): 

\be\label{taugravy}
\tau=10^{-65}  \,,\,
\tau=10^{-93}  \,,\, \tau=10^{94} \,,\, \tau=10^{307}\,,
\ee
to be compared with 
the corresponding results for the tunneling time obtained 
from the analysis performed in a flat spacetime background, 
where we have: 
\be\label{tauflat}
\tau=10^{-80} \,,\,
\tau=10^{-103} \,,\, \tau=10^{80} \,,\, \tau=10^{293}\,.
\ee

Eqs.\,(\ref{taugravy}) and (\ref{tauflat}) together
with Fig.\,\ref{fig:NPren} contain the main lesson of the 
present section. They definitely show that, even 
when gravity is included in the analysis, the presence of NP  at
high energy scales (even though much higher than the instability 
scale) can have an enormous impact on the vacuum lifetime. It is worth to remind 
here that when the calculation is performed in the curved  
spacetime background and the presence of high energy new physics
is not considered, the tunneling time is given by 
(\ref{gt}) ($\tau\sim 10^{661} T_U$), while from 
(\ref{taugravy}) we see that $\tau$ strongly depends on the
parameters of new physics, and can turn out to be even 
shorter than the age of the Universe. 

Moreover, by comparing (\ref{taugravy}) and (\ref{tauflat}) 
we see that gravity slightly pushes toward stabilization
(that is also what we observe for the case when new 
physics is not taken into account by comparing 
(\ref{tf}) with (\ref{gt})), but this is a ``tiny''
effect, that (as we have just seen) does not generate 
the claimed\,\cite{espinosa,ridolfi,espinosa2} 
disappearance of the new bounce solutions. 

Despite the fact that in our toy model NP  lives at 
very high energy scales, much higher than the instability
scale $\phi_{\rm inst}$, the expectation that the tunneling time 
should be insensitive to it, in other words that the result 
shown in \,(\ref{gt}) should not be modified by the 
presence of NP  at high energies, is not fulfilled. 
These results confirm the analysis of the previous section. 
Here, with the help of a 
fully renormalizable toy UV completion of the SM, we have shown 
that the EW vacuum lifetime strongly depends on NP  even 
if the latter lives at very high energy scales. 

These findings are at odds with a widely diffused expectation,  
based on a naive application of the decoupling argument, and
show that the fact that the vacuum stability condition 
depends on physics that lives at very high energy scales is not 
due to an illegitimate extrapolation of the theory beyond its 
validity, as it was previously thought\,\cite{Espinosa:2015qea}. 
On the contrary, it is an illegitimate application of the 
decoupling argument to a phenomenon to which it cannot be applied, 
namely the (non-perturbative) tunneling phenomenon, that 
leads to the expectation that physics at scales much higher 
than the instability scale $\phi_{\rm inst}$ should have no impact 
on the stability condition.

Before ending this section, we would like to stress once again
that with respect to the previous section, where NP
interactions were parametrized with the help of higher order 
non-renormalizable operators, here NP is 
given in terms of a fully renormalizable theory, thus showing 
that the effect that we present is a genuine physical effect 
and has nothing to do with the specific parametrization of NP. 

\section{Summary, conclusions, and outlook}

We studied the impact of very high energy NP  
(around the Planck scale $M_P$) on the stability 
condition of the EW vacuum by carrying the analysis in a curved 
spacetime background, i.e. by taking into account the presence 
of gravity. Despite the 
expectation\,\cite{espinosa,ridolfi, espinosa2} 
that strong gravity effects should act against the formation 
of new true vacuum bubbles, thus invalidating the results 
of previous analyses\,\cite{our1,our2,our3,our4} carried in a 
flat spacetime background, we found that these new solutions 
to the bounce equations persist even in the presence of gravity, 
and as in the case of the flat spacetime background analysis,
they can have an enormous impact on the EW vacuum lifetime. 
 
As in\,\cite{our1,our2,our3} we first performed the analysis   
by adding to the SM potential higher powers of the the Higgs 
field, more precisely terms as $\phi^6/M_P^2$ and 
$\phi^8/M_P^4$ (Section III) that are 
certainly generated in a quantum gravity context\,\cite{plefka}. 
Following\,\cite{our4} we then parametrized high energy new 
physics in a 
different manner, namely by adding to the SM potential a 
boson $S$ and a fermion $\psi$, with very large masses 
$M_S$ and $M_f$, coupled to the Higgs boson. As for the 
analysis carried in flat spacetime, in both 
cases we find that the presence of new physics
can have an enormous impact on the EW vacuum stability 
condition. 

These results definitely show that, 
irrespectively of the parametrization used to 
describe high energy new physics, it is not possible
to ignore its presence when the stability analysis is
performed. They are of the greatest 
importance for current studies and for model building 
of Beyond Standard Model physics, where we often 
have to take into account new physics at very high 
(Planckian and/or trans-Planckian) scales.

A question that is left open by the present analysis 
is the role that could be played by a non-minimal 
coupling of gravity to the Higgs boson. Work in this 
direction is in progress\,\cite{progress}.

\acknowledgements
This work is carried out within the INFN project 
QFT-HEP and is supported in part by the HARMONIA project 
under contract UMO-2015/18/M/ST2/00518 (2016-2019).
EB is also supported by the project ``Digitizing the 
universe: precision modelling for precision cosmology'', 
funded by the Italian Ministry of Education, University 
and Research (MIUR).

\newpage
\appendix

\section{Numerical computation of the bounce solution}

The search for the bounce solution is a boundary-value problem 
specified by the values of $\varphi'(0)$ and $\varphi(\infty)$. 
This can be turned into an 
initial-value problem using the shooting method, whereby
$\varphi(\infty)$ is replaced by $\varphi(0)$, and the 
appropriate value 
for the latter quantity is found iteratively,
solving Eq.\,(\ref{smeq}) (or its curved-space generalization)
for different initial values until the desired $\varphi(\infty)$ is obtained. 

As will be clear below, knowledge of the asymptotic behavior 
of $\varphi(x)$ for $x \to 0$ and $x \to \infty$ is 
a crucial ingredient for the efficiency of the shooting
algorithm.
To find the expected behavior of $\varphi(x)$ in the relevant regimes,
we begin by expanding $\varphi(x)$ around $x=0$:
\be 
\varphi(x)=B_0+B_2x^2+B_3x^3+...
\ee
where the linear term is missing due to the condition 
$\varphi'(0)=0$. Inserting this expansion in (\ref{smeq}), 
with $U(\varphi)$ given by\,(\ref{potenzialescalato})
we find that the coefficients of the odd-power terms 
vanish, $B_{2n+1}=0$, while all the coefficients of the even-power 
terms $B_{2n}$ are functions of the first coefficient
$B_0$ (called $B$ from now on). Truncating the expansion 
to the $x^2$ term:
\be \label{exp}
\varphi(x)=B+ \frac{B^3}{8} \left ( \lambda_*+ 
\frac \alpha 2 \ln B+ \alpha \ln^2 B + \beta \ln^3 B + \beta \ln^4 B 
\right )x^2 + ...
\ee
The coefficient of $x^2$ turns out to be negative, so near 
the origin the bounce profile behaves as an upside-down
parabola. 

As for the behavior of $\varphi(x)$ for 
$x \to \infty$, we note that $U(\varphi) \to 0$ for 
$x \to \infty$, and $\varphi(x) \to 0$ for $x \to \infty$. 
Asymptotically Eq.(\ref{smeq}) and the corresponding solution are
then:
\be \label{asymptotic}
\varphi''(x) + \frac 3 x \varphi'(x)=0 \quad 
\Rightarrow \quad \varphi(x)=\frac{A}{x^2}\,,
\ee
where $A$ is one of the integration constants, while 
the second additive integration constant vanishes due 
to the condition 
$\varphi(\infty)=0$. In other words, for the bounce solution  
$x^2 \varphi(x)$ has to reach a \emph{plateau} for 
$x \to \infty$.

\begin{figure}[!h]
          \includegraphics[width=0.9\textwidth]{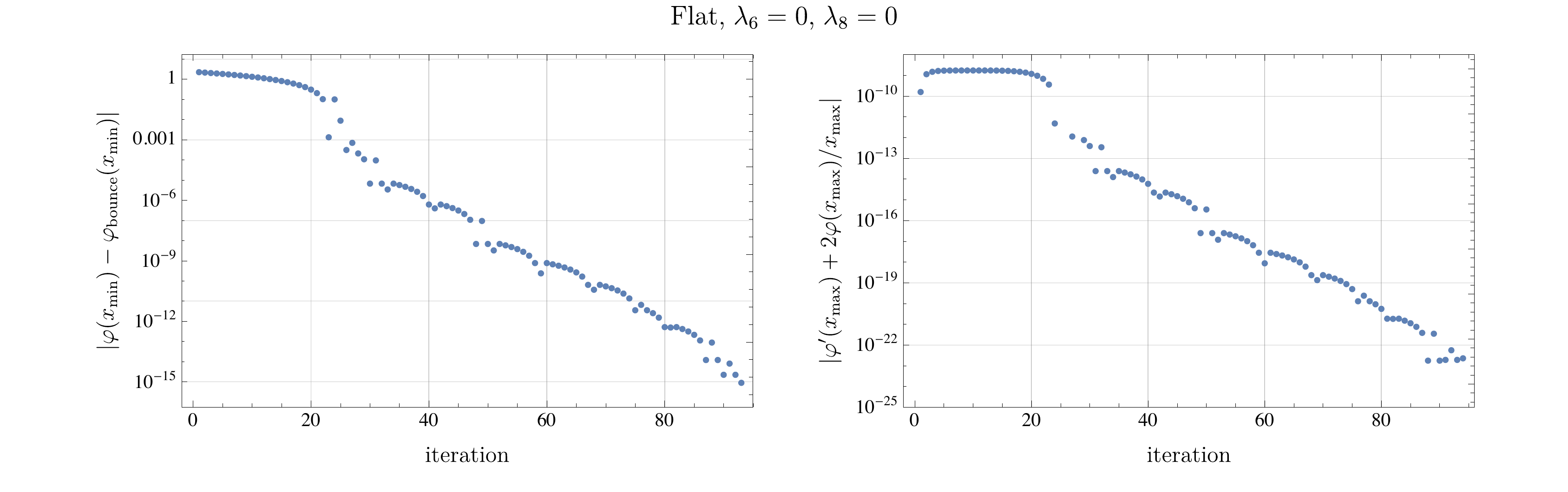}
          \includegraphics[width=0.9\textwidth]{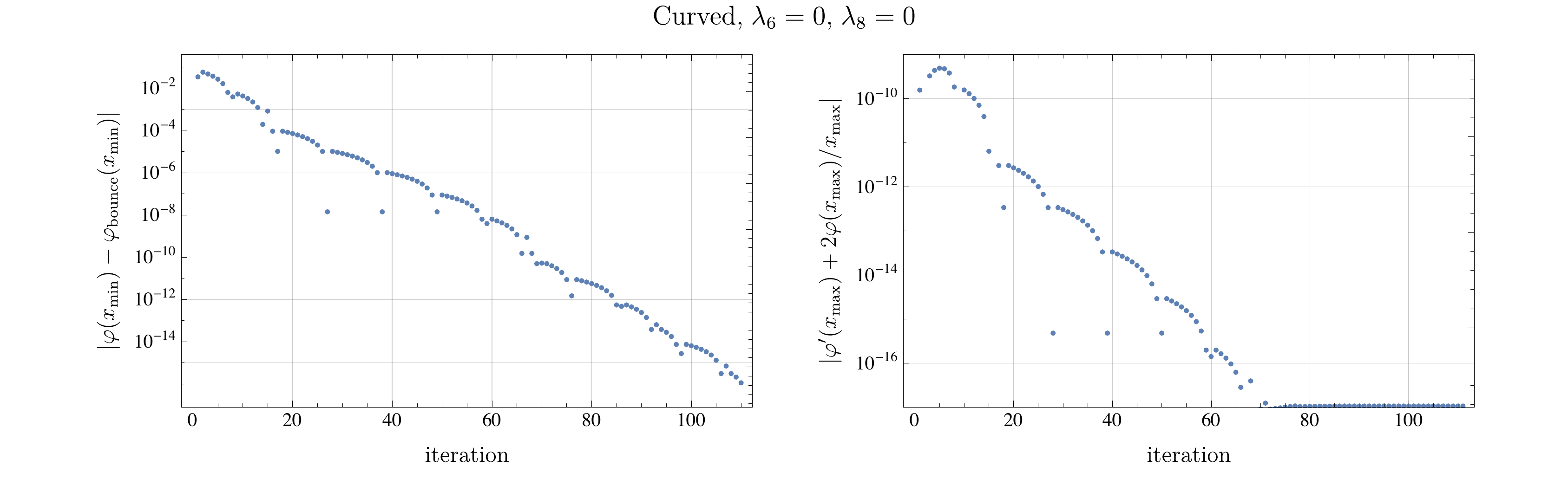}
          \includegraphics[width=0.9\textwidth]{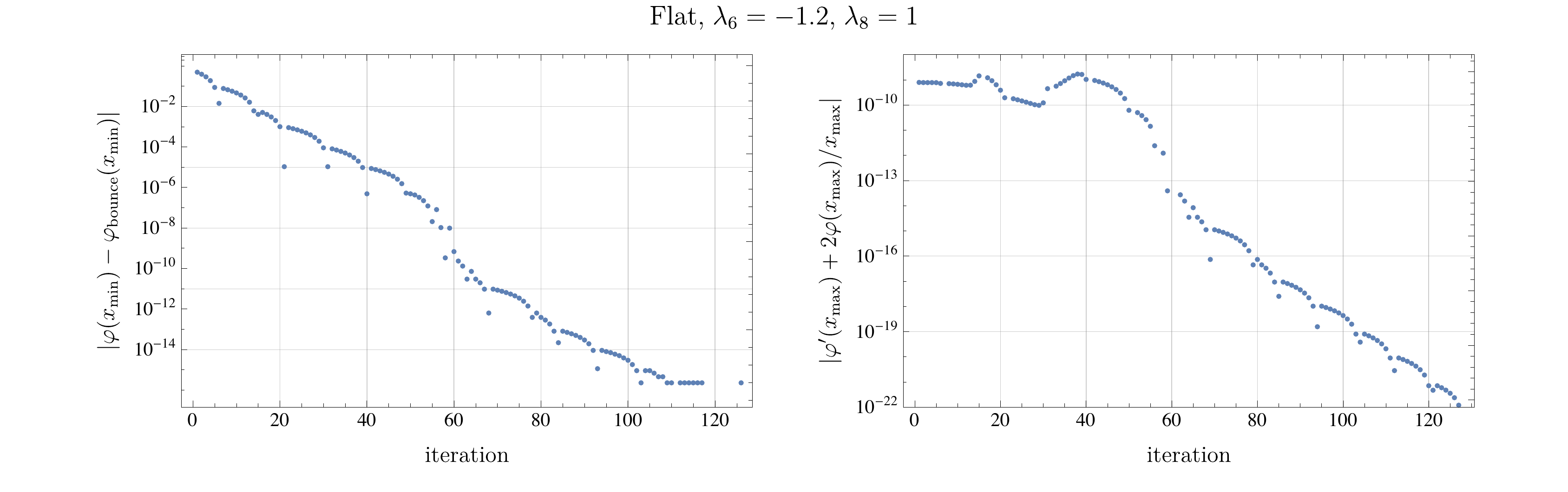}
          \includegraphics[width=0.9\textwidth]{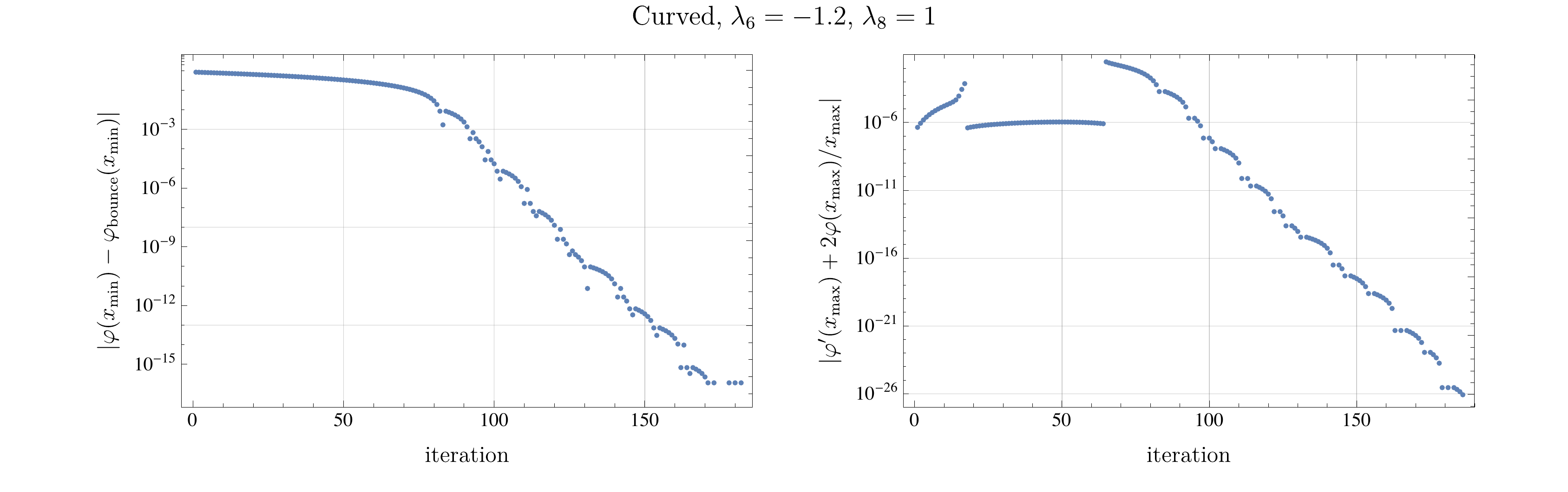}
                \caption{Convergence of $\varphi(x_{\rm min})$ to its
final value (left column),
                        as well as of $\varphi'(x_{\rm{max}})+2\varphi
(x_{\rm{max}})/x_{\rm{max}}$
                        to zero (right column), in the four cases (flat
and curved, with and without new physics)
                        discussed in the appendix.
                        \label{fig:algo}}
 \end{figure}

Numerically, we have implemented a fully adaptive algorithm 
designed to: (i) pick an initial guess ($\varphi(0)=B$, 
$\varphi'(0)=0$), (ii) integrate Eq.(\ref{smeq}) 
numerically while monitoring the behavior of $\varphi(x)$,
and (iii) iteratively restart the procedure with a suitably 
corrected $B$ until the condition $\varphi(\infty)=0$ is 
satisfied up to a prescribed tolerance.
In practice, the numerical integration is carried out in 
the range $[x_{\rm min}, x_{\rm max}]$, where we have chosen
$x_{\rm min} = 10^{-10}$ and $x_{\rm max} = 10^9$, so that 
the initial 
conditions for $\varphi$ and $\varphi'$  are given at $x=x_{\rm min}$ using Eq.\,(\ref{exp}). With this boundary conditions, we find a class of solutions $\varphi_B(x)$, parametrized by $B$. Following the \emph{overshoot-undershoot argument} of 
Coleman\,\cite{coleman}, we want to tune the parameter $B$ until we converge to the solution which reaches a plateau for $x \to \infty$.

We found that the characterization of the final state is of 
crucial importance to the effectiveness of the search.
In particular, introducing the reference point 
$x_{\rm ref}=x_{\rm max}-10^3$ and a tolerance $\epsilon=10^{-10}$, 
we found that the following three criteria are sufficient 
to lead the algorithm to the bounce solution in all cases 
(denoting $\tilde\varphi(x)=x^2\varphi(x)$):
\begin{enumerate}

\item If $\tilde \varphi(x_{\rm max})- \tilde \varphi(x_{\rm ref}) > 
\epsilon$, the scalar field has reversed its direction 
before reaching $\varphi=0$, and is returning towards its 
initial position. The initial guess for $\varphi(x_{\rm min})$ 
was therefore too low, and $B$ is correspondingly increased by a 
quantity $\delta$.

\item If $\tilde \varphi(x_{\rm max})- \tilde \varphi(x_{\rm ref}) <  
- \epsilon$, the scalar field is overshooting the top of the 
hill. The initial guess for $\varphi(x_{\rm min})$ 
was therefore too high, and $B$ is correspondingly decreased 
by a quantity $\delta$.

\item If $|\tilde \varphi(x_{\rm max})- \tilde \varphi(x_{\rm ref})| < 
\epsilon$, we consider that the plateau has been reached, i.e.
that the bounce solution is found within the required precision. 

\end{enumerate}

Furthermore, the algorithm stores a state consisting of both
$B$ and its value for the two preceding iterations. It is 
therefore possible to detect oscillations in $B$ and bisect 
(or otherwise decrease) $\delta$. We found that decreasing 
$\delta$ to $\delta/10$ each time $B$ changes trend leads to 
a particularly efficient search, that converges exponentially
to the solution, as will be illustrated below.

The inclusion of gravity does not modify our algorithm, as 
we found that, for $x \to \infty$, the areal radius goes 
as $a \sim x + c$ (see text), i.e. the curvature tends 
asymptotically to zero (the value of the constant $c$ 
is given by $a(x)-x$ when the plateau is reached).  

For this reason, the criteria for tuning the initial 
value of $\varphi$, introduced in the flat case, can also 
be used on a curved spacetime. Again, we implemented the 
boundary condition in $x_{\rm min}$:
			\be \label{nsgbc}
			\varphi(x_{\rm min})=B \qquad 
			\varphi'(x_{\rm min})=0 \qquad 
a(x_{\rm min})=\epsilon' \qquad a'(x_{\rm min})=1\,,
			\ee
with $\epsilon' \ll 1$ (clearly we cannot use $a(x_{\rm min})=0$ 
due to the factor $a^{-1}$ in (\ref{seqgrav})$_1$).
			
After solving the equations, the size ${\cal R}$ of the bounce 
solution is obtained and we can compute
the integral in Eq.(\ref{sgB}).
			
We go now back to the flat spacetime case and include the 
NP  terms of (\ref{pnp}), thus  
obtaining the (dimensionless) potential (\ref{newp}). 
The bonus for our analysis is that this potential does not modify the asymptotic behavior of the bounce solution $\varphi_b(x)$ for $x \to \infty$, as in this limit we still have $U(\varphi(x)) \to 0$. Therefore the inclusion of NP  does not lead to
any substantial change in our numerical method. The only modification with respect to the flat spacetime case concerns the expansion of the bounce solution around the origin $x=0$. In the flat case, by considering the integration range $[x_{\rm min},x_{\rm max}]$, we found that the initial values for solving Eq.\,(\ref{meq}) with the shooting method are obtained once we take the expansion (\ref{exp}) of $\varphi(x)$ and its first derivative at $x_{\rm min}$. 
Repeating the same analysis for the potential (\ref{newp}), we find that NP  simply leads to additional terms in these expressions. 
Thus, the new expansion for $\varphi(x)$ is given by (again up to 
$O(x^2)$):
\be \label{nbc1}
\varphi(x)=B+ \frac{B^3}{8} \left 
( \lambda_*+\lambda_6 B^2+\lambda_8 B^4+ 
\frac \alpha 2 \ln B+ \alpha \ln^2 B + \beta \ln^3 B + 
\beta \ln^4 B \right )x^2 +\cdots
\ee
which we use to set initial conditions for $\varphi(x_{\rm min})$ and $\varphi'(x_{\rm min})$.
An analogous approach is followed when we consider the
alternative parametrization of NP  given in 
Eq.\,(\ref{Vtot}).
Just like in the case without new physics, 
when we include gravity we use the initial values (\ref{nsgbc}), 
and the numerical integration of the equations of motion is 
reduced to the tuning of the parameter $B$.

Fig.\,\ref{fig:algo} illustrates the exponential convergence 
of our algorithm in the four cases presented in this appendix.

\end{document}